\documentclass[%
 reprint,
 amsmath,amssymb,
 aps,
]{revtex4-1}

\usepackage{graphicx}
\usepackage{dcolumn}
\usepackage{bm}

\usepackage{xcolor}

\usepackage{boites}
\usepackage{braket}

\begin{document}

\preprint{APS/123-QED}

\title{Conversion of Thermal Equilibrium States\\ 
into Superpositions of Macroscopically Distinct States}

\author{Mamiko Tatsuta}
 \email{mamiko@ASone.c.u-tokyo.ac.jp}
\author{Akira Shimizu}%
 \email{shmz@ASone.c.u-tokyo.ac.jp}
\affiliation{%
Department of Basic Science, University of Tokyo, 3-8-1 Komaba, Meguro, Tokyo 153-8902, Japan
}%




\date{\today}

\begin{abstract}
A 
simple procedure for obtaining  superpositions of macroscopically distinct states is proposed and analyzed. 
We find that 
a thermal equilibrium state 
can be converted into
such a state
when a single global measurement of 
a macroscopic observable, such as the total magnetization,
is made.
This method is valid for systems with macroscopic degrees of freedom 
and finite (including zero) 
temperature. 
The superposition state is obtained with a high (low) probability when the measurement is made with a high (low) resolution.
We find that this method is feasible in an experiment.
\end{abstract}

\pacs{Valid PACS appear here}
\maketitle


\section{introduction}
Superpositions of macroscopically distinct states 
have been attracting much attention 
\cite{schroedinger,leggett1980macroscopic,shmzmiyadera2002,yurke1986generating,mermin1990extreme,roy1991tests,kim1992schrodinger,q,mancini1997ponderomotive,bose1997preparation,p,zurek2003,sugitashmz2005,paternostro2011engineering,frowis2012measures,PhysRevLett.110.163602,morimaeshmz2006,akram2013entangled,vanner2013quantum,arndt2014,ghobadi2014optomechanical,morimae2010superposition,toth2014quantum,jeong2015characterizations,frowis2015linking,PhysRevB.93.195127,PhysRevA.94.012105,hoff2016measurement}. 
 Such superposition is craved in
quantum metrology \cite{ralph2002coherent,gerry2002nonlinear,munro2002weak,giovannetti2004quantum,giovannetti2011advances,joo2011quantum,facon2016sensitive},
quantum computation \cite{cochrane1999macroscopically,jeong2002efficient,ralph2003quantum,ukena2004appearance,cerf2007quantum,lund2008fault,speedup,heeres2016implementing,pfaff2016schrodinger},
quantum teleportation \cite{boschi1998experimental,van2001entangled,jeong2001quantum,neergaard2013quantum},
quantum repeaters \cite{sangouard2010quantum,brask2010hybrid,borregaard2012hybrid},
and fundamental tests of quantum mechanics \cite{greenberger1990bell,jeong2003quantum,stobinska2007violation}.
%
 Its experimental realizations have been reported
in various systems \cite{monroe1996schrodinger,brune1996,friedman2000quantum,leibfried2005creation,singlemode2007,deleglise2008,gao2010experimental,wineland2013nobel,haroche2013,vlastakis2013deterministically,kirchmair2013observation,wang2016schrodinger}.
  However,
 most of them
are limited to either  extremely low temperature, 
such as in the superconducting quantum interference devices \cite{friedman2000quantum},
or systems with small degrees of freedom, 
such as the single-mode photons \cite{singlemode2007}.
 These limitations were necessary
for a long coherence time
and good controllability.

For the coherence time, it seems possible to overcome the limitation;  
 recent experiments showed that
a long coherence time can be obtained even at room temperature, 
such as $\sim 2$ ms \cite{balasubramanian2009ultralong,ishikawa2012optical,maurer2012room}
in a negatively charged nitrogen-vacancy (NV$^-$) center 
in diamond \cite{nvcenter,doherty2013nitrogen,schirhagl2014nitrogen}.
A macroscopic system composed of 
a large number of such systems 
is expected to have both a long coherence time and large degrees of freedom,
even at moderate temperature such as $300$ K.
If superpositions of macroscopically distinct states are realized
in such a macroscopic system,
it is interesting not only from 
a fundamental viewpoint that macroscopic quantum coherence 
is proved possible at such high temperature, but also from a viewpoint on potential 
applications  
 operating 
at room temperature. 

For the controllability, however, 
a challenge seems to remain;
 a macroscopic system at finite temperature $T$ 
is expected to be in a thermal equilibrium state, i.e., 
 in the Gibbs state $\hat{\rho}_{\mathrm {eq}}$.
 Since $\hat{\rho}_{\mathrm {eq}}$ for $T>0$ 
is a classical mixture of an exponentially large number of quantum states,
it may seem very difficult to convert it into 
 superpositions of macroscopically distinct states.
 A successful example is 
the interference experiment using a C$_{60}$ molecule 
at $900$ K \cite{arndt1999wave,nairz2001diffraction}.
%
%
 In this molecule, however,
the center-of-mass motion,
which exhibits the interference,
 is 
completely decoupled from the internal motion,  which is responsible for the high temperature and high mixture.
 Such a perfect decoupling cannot be expected for general systems.
Then, 
 how can we convert  equilibrium states of general macroscopic systems 
into  such superpositions?

In this paper, we 
 propose a simple method for obtaining  superpositions of macroscopically distinct states from a thermal equilibrium state.
For systems with $N$ spins,
 for example,
we find that such a state is obtained through just a single global measurement of the total magnetization 
of an equilibrium state in a magnetic field. 
Although
the obtained state is a classical mixture of $\exp(\Theta(N))$ 
\footnote{
In addition to the standard notation
$O(N^k)$,
we use $\Theta(N^k)$:
For a function $f$ of $N$, we say 
$f = \Theta(N^k)$
if $f/N^k$ approaches a positive constant as $N \to \infty$.
If $f = \Theta(N^k)$ then $f=O(N^k)$,
but the inverse is not necessarily true.
}
states and has non-vanishing temperature, it contains superpositions of macroscopically distinct states with a significant ratio, and therefore has potential applications
like other cat states. 
 This method 
is applicable to a wide class of systems  including spin systems, atomic systems, quantum optical systems, and quantum dots.
 Through estimation using a state-of-the-art magnetometer and a system with long coherence time, we find our method feasible in an experiment.

\section{Recipe for conversion}
Our method is summarized as the following simple recipe.
 Although the recipe is applicable to various systems (see below), we explain it by taking, as an illustration, a quantum system composed of $N$ spins with $S=1/2$.
 First,
attach a heat bath of inverse temperature 
$\beta$ $(=1/T)$,
apply a magnetic field $\boldsymbol{h}=(h,0,0)$ parallel to the $x$ axis,
 and let the system equilibrate thermally. 
 Then the state of this system becomes
$ 
\hat \rho_{\mathrm{eq}}:=
e^{-\beta \hat{H}}
/Z_{\mathrm{eq}}.
$ 
Here, $\hat H$ is the Hamiltonian of the system in the presence of $\boldsymbol{h}$,
and 
$
Z_{\mathrm{eq}} :=
\mathrm{Tr} \big[ e^{-\beta \hat{H}} \big]
$.
We will discuss  details of $\hat H$ below.
Second,
measure  the $z$ component of the total  magnetization $\hat{M}_z=\sum_{i=1}^N\hat{\sigma}_z^{i}$
in such a way that the measurement operator \cite{nc,wiseman2010} is 
 the projection operator $\hat P_z$
onto the $\hat M_z=M$ subspace,
where $M$  denotes the outcome of the measurement.
 The case of a more general $\hat P_z$ will be discussed below.
If preferred, one may remove the heat bath and $\boldsymbol{h}$  just before the measurement (quickly, so as not to change the state).
Then, 
if the measurement is completed before the system decoheres,
the post-measurement state $\hat{\rho}_{\mathrm{post}}$ is given by
\begin{align}
\hat{\rho}_{\mathrm{post}}
=
\hat P_ze^{-\beta \hat{H}}\hat P_z
/
Z_{\mathrm{post}},
\label{rho}
\end{align}
where
$Z_{\mathrm{post}} :=
\mathrm{Tr}\big[\hat P_z e^{-\beta \hat{H}}\hat P_z\big]$ is the  `partition function,' which is not the ordinary 
 one
because we will show that $\hat \rho_{\mathrm{post}}$ is not a normal equilibrium state.
Note that 
$\big[ \hat{H},\hat{M}_z \big]\neq 0$
because of 
 the interaction $-h\hat M_x$ with the magnetic field.
 This noncommutativity makes
the  properties of  $\hat \rho_{\mathrm{post}}$ highly nontrivial. 
We will show that 
this state is a generalized cat state,
containing superpositions of macroscopically distinct state  with a significant ratio (precise definition  below).

\section{Generalized cat state}
Two states are macroscopically distinct if 
there is a macroscopic observable whose values are macroscopically 
distinct between them.
Using this general observation, several measures have been proposed 
to quantify superposition of macroscopically distinct states \cite{shmzmiyadera2002,q,frowis2012measures,park2016quantum},
which are motivated, e.g., by stability \cite{shmzmiyadera2002} and Fisher information \cite{frowis2012measures,frowis2015linking}.
For pure states, all these measures are found to be equivalent to 
that of Ref.~\cite{shmzmiyadera2002}.
When a pure state is a superposition of macroscopically distinct states
according to this established measure, we call it a {\em pure cat state}.

As we will show below, 
$\hat \rho_{\mathrm{post}}$ is a mixture of $\exp(\Theta(N))$ states
for $T>0$.
In order to detect superpositions of macroscopically distinct states
in such a state, 
 we use the index $q$ \cite{q,frowis2012measures,morimae2010superposition,toth2014quantum,jeong2015characterizations,frowis2015linking,PhysRevB.93.195127} 
 (briefly reviewed in Appendix \ref{sec:sup:q})
because it correctly detects pure cat states 
(if contained) 
in any mixed state.
It is sufficient for the present purpose
to introduce  $q$ as a real number satisfying
\begin{align}
\max_{\hat A, \hat \eta}\mathrm{Tr}\big[\hat \rho \hat C_{\hat A \hat \eta}\big]
=\Theta (N^q)\label{qdef}
\end{align}
for a general state $\hat \rho$. 
Here, 
$\hat A$ is an additive observable,  $\hat \eta$ is a projection operator,
and 
$
\hat C_{\hat A \hat \eta}:=\big[\hat A,\big[\hat A, \hat \eta \big]\big].
$
It is easily shown that  $q \leq 2$.
Supported by reasonable observations
(see also Appendix \ref{sec:sup:q} for details),
Ref.~\cite{q}   showed that if some $\hat A$ and $\hat \eta$ assure $q=2$ for a quantum state, 
 then the state contains  pure cat states, i.e., superpositions of states whose values of $\hat A$ differ from each other by $\Theta (N)$, with a significant ratio.
We call such a state {\em a generalized cat state of $\hat A$}.
By ``general,'' we mean a state with $q=2$ is not necessarily a pure state or a superposition of only two states like Schr\"odinger's cat state.
 For example, 
 the index $q$ correctly identifies the mixed state (which beats the standard quantum limit using quantum superposition)
of Ref.~\cite{Jones1166}  as a generalized cat state.
When necessary, one can also use the value of 
$\braket{\hat C_{\hat A \hat \eta}}$ 
as a quantitative measure (Appendix \ref{sec:quantify}).

\section{Example of free spins}
Though we will show below that our recipe is applicable to varieties of Hamiltonians,
 we start with the case of 
free spins
 to grasp the idea.
The Hamiltonian, after applying the magnetic field, is
\begin{equation}
\hat{H}^0:=-h\sum_{i=1}^N\hat{\sigma}_x^{i}=-h\hat{M}_x,\label{h0}
\end{equation}
where the superscript ``0'' denotes the absence of interactions between spins. 
The system is left to equilibrate in a heat bath of inverse temperature $\beta$. 
The state of the system becomes
$
\hat \rho^0_{\mathrm{eq}}:=e^{-\beta \hat{H}^0}/Z^0_{\mathrm{eq}},
$
 where
\begin{align}
Z^0_{\mathrm{eq}}:=\mathrm{Tr}\big[e^{-\beta \hat{H}^0}\big]
=2^N\cosh^N (\beta h).
\label{z0eq}
\end{align}
For this $\hat \rho^0_{\mathrm{eq}}$, $\hat M_z$ is measured, and the outcome $M$ is obtained.
Then the post-measurement state $\hat \rho_{\mathrm{post}}^0$ is given by
Eq.~(\ref{rho}) with $\hat{H} = \hat{H}_0$,
and $Z_{\mathrm{post}}$ (denoted as $Z^0_{\mathrm{post}}$) by
\begin{equation}
Z^0_{\mathrm{post}} :=
\mathrm{Tr}\big[\hat P_ze^{-\beta
    \hat{H}^0}\hat P_z\big]
= \binom{N}{\frac{N+M}{2}}\cosh^N( \beta h)\label{z-0}.
\end{equation}

While $\braket{\hat{M}_x}_{\mathrm{eq}}= \Theta(N)$,
we can easily show that 
$\braket{\hat{M}_x}_{\mathrm{post}}=\braket{\hat{M}_y}_{\mathrm{post}}=0$,
where, throughout this paper, $\braket{\cdot}_{\mathrm{eq}}$ and $\braket{\cdot}_{\mathrm{post}}$ denote the expectation values in the pre-measurement equilibrium state and the post-measurement state, respectively.
Therefore, the uncertainty relation 
$\delta M_y \delta M_z 
\geq |\braket{\hat{M}_x}|$,
used in discussions on spin squeezing \cite{kitagawa1993squeezed,julsgaard2001experimental,chen2011conditional,PhysRevLett.110.163602,PhysRevA.94.012105},
does not tell anything nontrivial 
for $\hat \rho_{\mathrm{post}}^0$.

To show $q=2$ for $\hat \rho_{\mathrm{post}}^0$, it is sufficient by definition to show $\braket{\hat C_{\hat A \hat \eta}}_{\mathrm{post}}=\Theta (N^2)$ for {\it some} $\hat A$ and $\hat \eta$.
Now, let us take
$
\hat{A}=\hat{M}_x
$
and
$
\hat{\eta}=\hat P_z.
$
 Then, 
 we have 
\begin{align}
\braket{\hat C_{\hat M_x \hat P_z}}_{\mathrm{post}} 
=2N+(N^2-M^2)\tanh^2(\beta h).
\label{cindep}
\end{align}
The  factor $N^2-M^2$ is $\Theta (N^2)$ 
 except when $M=\pm N +o(N)$, which is the case very unlikely to happen for the following reason.
From (\ref{z0eq}) and (\ref{z-0}),
the probability of obtaining $M$ as the outcome of the $\hat M_z$ measurement is calculated as 
$
\mathrm{Pr}\big[\hat M_z=M\big]
=Z^0_{\mathrm{post}}/Z^0_{\mathrm{eq}}
=\binom{N}{(N+M)/2}/2^N.
$
 Therefore,
$
\mathrm{Pr}\big[\hat M_z=\pm N+o(N)\big]
= N^{o(N)}/[o(N)]!2^N,
$
which is exponentially small in $N$.
Hence, we hereafter exclude the case $M=\pm N +o(N)$ and assume
\begin{align}
N^2-M^2=\Theta (N^2),\,\mathrm{i.e.},\,N-|M|=\Theta (N).\label{n2-m2}
\end{align}
Then,
since $\tanh^2(\beta h)=\Theta(1)$ for any $|\beta h|=\Theta(1)$, 
we have $\braket{\hat C_{\hat M_x \hat P_z}}_{\mathrm{post}}=\Theta (N^2)$,
and thus $q=2$.
 Therefore, an equilibrium state of a spin system with the Hamiltonian $\hat H^0=-h\hat M_x$ at any finite temperature 
 can be converted into a generalized cat state of $\hat M_x$ by measuring $\hat M_z$   just once.

\section{Features of the post-measurement state}
We note that 
the purity of 
$\hat \rho_{\mathrm{post}}^0$ is
$
\mathrm{Tr}[(\hat \rho_{\mathrm{post}}^0)^2]
\leq
1/\exp(\Theta(N))
$ when $|\beta h|=\Theta(1)$.
This low purity is due to two facts:
The purity of the pre-measurement state $\hat \rho_{\mathrm{eq}}^0$ is also $1/\exp(\Theta(N))$,  
and 
the subspace onto which $\hat P_z$ projects  has a dimension of $\exp(\Theta(N))$,
i.e., 
$\mathrm{Tr}[\hat P_z]=\binom{N}{(N+M)/2}$.
Therefore, 
 $\hat \rho_{\mathrm{post}}^0$ is a mixture of $\exp(\Theta(N))$ states when $|\beta h|=\Theta(1)$
despite containing  superpositions of macroscopically distinct states. 
This does not mean that the ratio of pure cat states  
(to non-cat states) 
is exponentially small,  according to the definition of the index $q$.
In fact, 
the factor $\tanh^2(\beta h)$ in Eq.~(\ref{cindep}) 
quantifies the contribution from pure cat states; 
it quantifies 
the ratio of pure cat states and 
how distinct are the values of $\hat{M}_x$
between the states that are superposed (Appendix \ref{sec:quantify}). 
 At intermediate temperatures where $\beta h =\Theta(1)$, 
there is $\tanh^2(\beta h)=\Theta(1)$ of contribution from pure cat states.
%


We  can show 
that $\braket{\hat H^0}_{\mathrm{post}}=0$ and 
$\braket{(\Delta \hat H^0)^2}_{\mathrm{post}}= \Theta (N^2)$ 
for  the generalized cat state $\hat \rho_{\mathrm{post}}^0$,
where $\Delta \hat H^0 := \hat H^0 - \braket{\hat H^0}_{\mathrm{post}}$.
Hence, 
if the energy  of this state is measured,
the outcome $E$ will  vary by $\Theta (N)$ 
from run to run.
 Let us define `temperature' of 
a nonequilibrium state (like the generalized cat state)
 as the temperature of the equilibrium state of the same energy.
[This definition is reasonable 
from  an operational viewpoint, 
as discussed in Appendix \ref{sec:temp}.]
 Then, 
 the `temperature' of $\hat \rho_{\mathrm{post}}^0$
varies 
from run to run. 
 Their values are non-zero except when $E=\pm h N$, the highest and lowest energies.
 In this sense,
the obtained generalized cat state has non-vanishing temperature.


\section{Extention of projection operator onto a finite interval}\label{debusec}
So far, the projection operator $\hat P_z$ which we used projects states onto the subspace of one exact eigenvalue $M$ of 
 $\hat M_z$. 
Although such a measurement 
 seems feasible with the present-day technologies (as discussed below),
there may be the cases where it is challenging to, for example, distinguish a state of $\hat M_z=0$ from a state of $\hat M_z=2$ because of a low resolution. 
 To model such a general case, 
we specifically consider the case with the projection operator $\hat P'_z$ onto the subspace corresponding to a finite interval $M_-\leq \hat M_z \leq M_+$.
 We here show that even with a low-resolution, one can in principle obtain a generalized cat state.
As an illustration, we study the system with $N$ free spins.
We assume $|M_-|<M_+$ without loss of generality,
 and 
evaluate the index $q$ of 
 the post-measurement state $\hat P'_ze^{-\beta \hat{H}^0}\hat P'_z/Z'^0_{\mathrm{post}}$.
 Here, $Z'^0_{\mathrm{post}}:=\mathrm{Tr}\big[\hat P'_ze^{-\beta \hat{H}^0}\hat P'_z\big]$ is the `partition function,' 
calculated as 
$
Z'^0_{\mathrm{post}} 
=
\sum_{k=0}^{(M_+-M_-)/2}\binom{N}{(N+M_-)/2+k}
\cosh^N\left(\beta h\right).$
Then,
using a function 
$
I(N,M_+,M_-)
$
given in Appendix \ref{sec:appdebu}
that satisfies $0 \leq I \leq 1$,
we have
\begin{align}
\!\! 
\braket{\hat C_{\hat M_x \hat P'_z}}_{\mathrm{post}}
\!\! 
=
N^2\tanh^2(\beta h)
I(N,M_+,M_-)
+O(N).
\label{debuc}
\end{align}
This becomes $\Theta (N^2)$ when 
$I(N,M_+,M_-)=\Theta(1)$.
After some algebra, we find that, when 
$M_+-M_-= \Theta(1)$,
the post-measurement state is a generalized cat state as long as $N-|M_-|=\Theta (N)$.
On the other hand, when $M_+-M_-> \Theta(1)$ \footnote{
By $f(N)>\Theta(N^k)$, we mean that for all $G>0$, there exists $N_G$ such that $f(N)/N^k>G$ for $\forall N \geq N_G$.
},
the post-measurement state is a generalized cat state only when $M_-= \Theta (N)$ and $N-|M_-|=\Theta (N)$.

There is a trade-off between the resolution, $M_+-M_-$, and the success probability.
If $M_+-M_-$ is as small as $\Theta (1)$, 
one can obtain a generalized cat state through our recipe with the success probability of almost $100\%$.
 Easier to realize is 
a measurement with a lower resolution of $M_+-M_->\Theta (1)$ [such as $\Theta (\sqrt N)$]. 
In this case, as described above,
$M_-$ has to be $\Theta (N)$ for obtaining a generalized cat state.
That is, 
the measured $\hat M_z$ has to be $\Theta (N)$.
However, the probability of such a case is
 exponentially small because
$
\mathrm{Pr}\big[\Theta (N)=M_- \leq \hat M_z \leq M_+\big]
=Z'^0_{\mathrm{post}}/Z^0_{\mathrm{eq}}
= e^{-\Theta (N)}.
$
%
 Thus, when $M_+-M_->\Theta (1)$, experiments should be run many times in order to obtain a generalized cat state.
When designing experiments, these conditions should be taken into account according to one's purpose.

\section{Generallization of systems and initial states}
 Up to this point, we have  assumed spin-$1/2$ systems and the canonical Gibbs states as the pre-measurement states.
Actually, any two-level system can be mapped to a spin-$1/2$ system. 
Thus our discussion is already applicable to other physical systems such as two-level atoms by mapping observables such as $\hat M_x$ and $\hat M_z$ appropriately. 
We here add even more general discussion by providing two conditions. 
 At the same time, we show that the initial states need not  be the Gibbs states.

Consider a macroscopic quantum system, which is not necessarily a spin system, in some state $\hat \rho_{\mathrm{pre}}$,
which is taken as the pre-measurement state. 
Let $\hat A$ and $\hat B$ be additive operators 
of the system such that
\begin{equation}
\hat P_{b} \hat A\ket{b,\xi}=0 \label{cond1}
\end{equation}
for all eigenstates $\ket{b,\xi}$ of $\hat B$, where $\hat{B}\ket{b,\xi}=b\ket{b,\xi}$ and $\xi$ labels degenerate eigenstates,
and $\hat P_{b}$ is the projection operator onto the $\hat B=b$ subspace.
When $\hat B$ of $\hat \rho_{\mathrm{pre}}$ is measured,
the post-measurement state $\hat \rho_{\mathrm{post}}$ is 
$\hat P_{b} \hat \rho_{\mathrm{pre}} \hat P_{b}/\mathrm{Tr}\big[\hat P_{b} \hat \rho_{\mathrm{pre}} \hat P_{b}\big]$.
We then obtain $\braket{\hat A}_{\mathrm{post}}=0$
and, by taking $\hat \eta =\hat P_b$,
$\braket{\hat C_{\hat A \hat P_b}}_{\mathrm{post}}=2\mathrm{Tr}\big[\hat P_{b} \hat \rho_{\mathrm{pre}} \hat P_{b}\hat A^2\big]/\mathrm{Tr}\big[\hat P_{b} \hat \rho_{\mathrm{pre}} \hat P_{b}\big]$.
Thus, if the set $\{\hat A, \,\hat B,\,\hat \rho_{\mathrm{pre}} \}$ satisfies
\begin{equation}
{\mathrm{Tr}\big[
\hat P_{b} \hat \rho_{\mathrm{pre}} \hat P_{b}\hat A^2
\big]}/{\mathrm{Tr}\big[
\hat P_{b} \hat \rho_{\mathrm{pre}} \hat P_{b}
\big]}
=
\Theta (N^2), \label{cond2}
\end{equation}
then $\hat \rho_{\mathrm{post}}$ 
is a generalized cat state.
 [In the case of Eq.~(\ref{cindep}), 
for example, the set of
$\{\hat M_x, \,\hat M_z,\,\hat \rho_{\mathrm{eq}} \}$ corresponds to 
$\{\hat A, \,\hat B,\,\hat \rho_{\mathrm{pre}} \}$.]
Since this  result is applicable to any systems including quantum optical systems, atomic systems, and quantum dots, our recipe can be carried out in a wide class of physical systems.

The  sufficient conditions (\ref{cond1}) and (\ref{cond2})
tell us
that $\hat \rho_{\mathrm{pre}}$ 
is not required to be 
 the canonical Gibbs state $\hat{\rho}_{\mathrm{eq}}$.
 For example, in spin systems 
(or systems that can be mapped to spin systems),
the pre-measurement state $\hat \rho_{\mathrm{pre}}$ 
may be {\em arbitrary} if it has a macroscopic value of $\hat M_x$, 
i.e.,  if
$\braket{\hat M_x}_{\mathrm{pre}}=\Theta(N)$,
because then the conditions are satisfied
 by the set of 
$\{ \hat{M}_x$ or $\hat{M}_y, \hat{M}_z, \hat \rho_{\mathrm{pre}} \}$  with a non-vanishing probability (Appendix \ref{sec:koashi}).
 This sufficient condition 
indicates
 that details of the system does not matter to our recipe.
For example, our recipe is applicable even to systems with interactions whether the interactions are short-range or long-range. A detailed example of the $XYZ$ model  and a discussion using symmetries are in Appendix \ref{sec:xyz} and \ref{sp:sym}. Appendix \ref{sec:appdebu} also suggests that the discussions of \ref{debusec}
 hold the same for interacting spins.

\section{Feasibility}
To carry out our recipe, 
we need a  spin system which does not decohere during the $\hat M_z$ measurement.
Long  coherence times are realized in various systems such as those with ultracold atoms \cite{ultracoldatom1,ultracoldatom2}
and
circuit QED  systems \cite{nakamura1999coherent,nakamura}.
 Among them, we here investigate the feasibility 
for the NV$^-$ centers 
\cite{nvcenter,balasubramanian2009ultralong,ishikawa2012optical,maurer2012room,doherty2013nitrogen},
in which 
a spin 
has a long coherence time   such as $470$ $\mu$s (or $2$ ms using a spin echo)
even at room temperature \cite{balasubramanian2009ultralong,ishikawa2012optical,maurer2012room}.
(Since $S=1$ is equivalent to two of $S=1/2$ spins, we can apply our recipe to the NV$^-$ centers which have $S=1$.)
Following our recipe,
we suppose that a system composed of  $N$ NV$^-$ centers  is left to equilibrate 
in the presence of a magnetic field.
To obtain a generalized cat state with a high probability,
we need to measure $\hat M_z$ with  the resolution of $\Theta(1)$  within $\tau_{\mathrm{coh}}$, the coherence time of the system.
Since the coherence of an $N$-spin system is lost when just one spin decoheres, 
$\tau_{\mathrm{coh}}$ is shorter than the coherence time $\tau$ of a single spin
(for more details, see Appendix \ref{sec:appfeasi}).
Here we assume a typical case $\tau_{\mathrm{coh}}=\tau/N$  to discuss the feasibility.
Fortunately,
 state-of-the-art magnetometers
\cite{toida2016electron,happer1973spin,allred2002high,savukov2005effects,dang2010ultrahigh},
such as 
the one based on optically pumped potassium atoms operating in a spin-exchange relaxation free (SERF) regime \cite{happer1973spin,allred2002high,savukov2005effects,dang2010ultrahigh},
are estimated to be sensitive enough: $160$ aT/$\sqrt{\mathrm{Hz}}$ \cite{dang2010ultrahigh}, for example.
 Since one spin creates a magnetic field ${\mu_{B} \mu_0}/{2\pi r^3}$  at distance $r$, 
measurement of $\Theta (1)$ resolution can be performed 
 within $\tau_{\mathrm{coh}}/10$ even at room temperature 
 when $N\lesssim 10^4$
and  $r\simeq 1$ $\mu$m.
In particular,
when $N=10^2$ and $r\simeq 1$ $\mu$m,
a generalized cat state is obtained with a measurement time $\ll \tau_{\mathrm{coh}}$ even at room temperature. 
Hence the obtained generalized cat state  survives for most of $\tau_{\mathrm{coh}}=4.7$ $\mu$s after conversion.

More generally, temperature $T$ of a system is ``high" if $k_{\rm B} T$ is larger than any relevant energy scale of the system even if $T$ is much lower than room temperature.
For free spins, for example, $T$ is high if $\beta h \lesssim 1$.
In this sense, 
a generalized cat state with a longer coherence time at sufficiently high temperature may be realized
in other systems such as electron spins in donors in high-purity Si \cite{eec6e647c2824bbab76a44bb03dd0eeb}.

 
 We can also show that within the system's coherence time, 
the post-measurement state continue being a generalized cat state while evolving with time (Appendix \ref{sec:time}).

\section{Verification of the conversion}
We also discuss how to verify the success of the creation of a  generalized cat state in experiments.
%
One way, which 
seems most practical, is to see the enhancement of performances of applications, 
such as the increase of the sensitivity in the application discussed above. 

Another way is to investigate the state itself as follows.
If 
$T \ll h$,
the pre-measurement state is the ground state, a pure state.
In this case, the post-measurement state will also be a pure state. 
 For a pure state $\ket{\psi}$, 
it was shown that if
$
\max_{\hat A}\braket{\psi|
(\Delta\hat A)^2
|\psi}=\Theta (N^2)
$,
where $\Delta \hat A := \hat A - \braket{\psi|\hat A|\psi}$,
 then $\ket{\psi}$ is a pure cat state \cite{shmzmiyadera2002,p,speedup,frowis2012measures,toth2014quantum,jeong2015characterizations,frowis2015linking,PhysRevB.93.195127} (see Appendix \ref{sec:sup:p} for a brief review).
For the free spins, for example,
 the success can be verified by measuring 
 $\hat M_x$, and 
 thereby confirming 
that the measured
$\braket{\hat M_x^2}_{\mathrm{post}}$ 
agrees with the theoretical result, calculated as
$N+(N^2-M^2)/2$ for $T=0$.
If 
$T \gg h$ and thus
the post-measurement state is a mixed state, investigation of the state is very difficult in general. 
For example, state tomography requires an exponentially large number of procedures, addressing  individual spins.
However,  fortunately, 
one can verify the success of conversion within 
the number of procedures that is
polynomial in $N$ for our case.
It is done by measuring $\hat C_{\hat M_x \hat P_z}$ and comparing the result with the theoretical one (\ref{cindep}).
 Since any observable of a spin system is a function of 
the Pauli operators, 
$\hat C_{\hat M_x \hat P_z}$ 
can be measured, by  addressing  individual spins (see Appendix\ref{sec:veri} for details). 

\section{Advantages and applications}
Here we summarize advantages of our recipe. 
(i) The temperature of the system is arbitrary, i.e.\ 
ultra low temperature is not required. 
(ii) The procedure is simple: just one global measurement.
(iii) Precise control of the pre-measurement state is unnecessary since 
 Eqs.~(\ref{cond1}) and (\ref{cond2}), or, more simply,  
$\braket{\hat M_x}_{\mathrm{pre}}=\Theta(N)$, is sufficient.
 (iv) It is applicable to many physical systems,  as discussed above.

 Our generalized cat state 
can be used for various applications like other cat states.
For example, 
as shown in Refs.~\cite{Jones1166,matsuzaki2011magnetic},
pure cat states and their mixture
improve the sensitivity of
magnetometry,
beating 
the standard quantum limit
 by a factor of $N^{1/4}$  even under the effect of decoherence.
 [Otherwise, enhancement is $N^{1/2}$, reaching the Heisenberg limit.]
 For the case of our generalized cat state,
we must take account of
the factor $\tanh^2(\beta h)$, 
which quantifies contribution from pure cat states,
as discussed above.
%
%
 However, since $\tanh^2(\beta h)$ is independent of $N$, the overall sensitivity beats the standard quantum limit by a factor proportional to $N^{1/4}$. 

\section{summary}
In summary, we proposed a simple method for obtaining a generalized cat state 
 through one global measurement. 
For spin systems, 
 an equilibrium state with a macroscopic value of $\hat M_x$
can be converted into
a generalized cat state  as the post-measurement state 
of the measurement that projects 
the equilibrium state 
onto the subspace
corresponding to an interval 
$M_-\leq \hat M_z \leq M_+$.
 The success probability is 
high (low) when $M_+-M_-=\Theta (1)$ ($M_+-M_->\Theta (1)$).
 We showed two loose conditions that show the applicability of our recipe to various systems such as free spins, interacting spins,  and more general systems including quantum optical systems and atomic systems.
We also estimated that the method is feasible when the SERF magnetometer is used for measuring the magnetization of the NV$^-$ centers in diamond. 


\begin{acknowledgments}
We thank M. Koashi
 for suggesting that 
$\braket{\hat M_x}_{\mathrm{pre}}=\Theta(N)$
is a sufficient condition.
We also thank
M. Ueda,  Y. Matsuzaki,
K. Fujii, H. Tasaki, N. Shiraishi, Y. Nakamura, 
J. Hayase, Y. Oono, H. Hakoshima and R. Hamazaki
for discussions.
This work was supported by The Japan Society
for the Promotion of Science, KAKENHI No. 26287085 and No. 15H05700.
M.T. was supported by the
Japan Society for the Promotion of Science through Program
for Leading Graduate Schools (ALPS).
\end{acknowledgments}

\appendix
\begin{widetext}
\section{Temperature of post-measurement state}\label{sec:temp}

Since the post-measurement state $\hat{\rho}_{\mathrm{post}}$ 
(which is denoted as $\hat{\rho}^0_{\mathrm{post}}$ for free spins)
is not an ordinary 
equilibrium state, its temperature cannot be defined trivially.
Therefore, 
it seems legitimate to define the `temperature' operationally,
i.e., to define it as 
the outcome that is obtained when the temperature of $\hat{\rho}_{\mathrm{post}}$
is really measured.

There are two typical methods of measuring temperature.
One is to use a thermometer which is much smaller than the system
so that it does not alter the temperature of $\hat{\rho}_{\mathrm{post}}$ of 
the system.
In our case, however, 
it is not clear whether the composite system 
composed of the system and 
such a small thermometer reaches equilibrium, 
because $\hat{\rho}_{\mathrm{post}}$ is far from equilibrium and we allow 
the system Hamiltonian to be integrable (such as free spins).
Therefore, we employ the other typical method, 
which is to use heat baths 
that are much larger than the system. 

Suppose that we have 
heat baths with various temperature,
and 
many copies of the system 
in the same state $\hat{\rho}_{\mathrm{post}}$.
Then, suppose that one heat bath 
 is attached to one copy of the system.
After some amount of energy flows between the bath and the system, 
this composite system will reach equilibrium.
If the net energy flow is zero, the temperature of the heat bath 
may be identified with the measured temperature of the system.
If, on the other hand, the net energy flow 
is nonzero, 
retry this experiment using another heat bath 
of another temperature, and another copy of the system.
This seems a reasonable and widely applicable method of measuring temperature.
%

When this method is used to measure the temperature $T_0$ of 
the post-measurement state $\hat{\rho}_0$ of free spins,
the measured temperature varies from measurement to measurement
because $\langle (\Delta \hat{H}_0)^2 \rangle = \Theta(N^2)$
in this state.
Since $\langle \hat{H}_0 \rangle = 0$ is $\Theta(N)$ 
larger than the ground energy $-h N$,
the average of the inverse temperature $1/T_0$,
over many runs of measurements, is finite.
In this sense, $\hat{\rho}_0$ has non-vanishing temperature.
In a similar manner, 
we can show that 
the post-measurement state $\hat{\rho}_{\mathrm{post}}$ of interacting spins
also has non-vanishing temperature.
That is, we can obtain by our recipe 
a generalized cat state of
non-vanishing temperature.

\section{Index $p$ for pure states}
\label{sec:sup:p}

In this section, we review index $p$, 
which for pure states detects superposition of macroscopically distinct states \cite{shmzmiyadera2002,p,sugitashmz2005,ukena2004appearance,morimaeshmz2006,speedup,q}. 
Although such states are called ``anomalously fluctuating states''
in Ref.~\cite{shmzmiyadera2002} and ``macroscopically entangled states'' 
in Refs.~\cite{p,ukena2004appearance,q},
we here call them  {\em pure cat states}
(or, {\em generalized cat states} for mixed states)
to be more comprehensible.

Although we have mainly used $q$ in the text, 
it seems {\em necessary to 
understand $p$ for understanding $q$.}
We therefore review $p$ first in this section, 
and then $q$ in Appendix \ref{sec:sup:q}.

We limit ourselves to pure states that are macroscopically uniform 
spatially. (The case of non-uniform states was formulated in Ref.~\cite{speedup}.)

%
%



\subsection{Motivation}

It is not trivial 
to define superposition of macroscopically distinct states.
For example, while the cat state
\begin{equation}
| \mbox{cat} + \rangle
\equiv
{1 \over \sqrt{2}} \ 
| 0 0 0 \cdots 0 \rangle
+
{1 \over \sqrt{2}} \ 
| 1 1 1 \cdots 1 \rangle
\end{equation}
is obviously such a superposition, how about the following states?
\begin{align}
| \psi_1 \rangle 
&\equiv
\sqrt{1- {1 \over N}} \
|000 \cdots 0 \rangle
+
\sqrt{{1 \over N}} \
|111 \cdots 1 \rangle,
\\
| \psi_2 \rangle 
&\equiv
{
|000 \cdots 0 \rangle
+
|100 \cdots 0 \rangle
+
|110 \cdots 0 \rangle
+
\cdots
+
|111 \cdots 1 \rangle
\over \sqrt{N+1}}.
\end{align}
%
In order to 
identify superposition of macroscopically distinct states
unambiguously, 
a reasonable index was proposed in Refs.~\cite{shmzmiyadera2002,p}, as follows.

\subsection{Macroscopically distinct states}

We start with defining `macroscopically distinct states.' 
Such states should be defined 
as those between which some macroscopic observable takes distinct values.
But, what is a `macroscopic observable'?
According to thermodynamics and statistical mechanics, 
macroscopic observables should be additive observables.
Here, we say $\hat{A}$ is an {\em additive observable} if it is the sum of 
local observables $\hat{a}(\bm{r})$,
\begin{equation}
\hat{A} = \sum_{\bm{r}} \hat{a}(\bm{r}),
\end{equation}
where the sum is taken over the whole system.
We assume, for simplicity, that\footnote{
 In addition to the standard notation
$O(N^k)$,
we use $\Theta(N^k)$:
For a function $f$ of $N$, we say 
$f = \Theta(N^k)$
if $f/N^k$ approaches a positive constant as $N \to \infty$.
If $f = \Theta(N^k)$ then $f=O(N^k)$,
but the inverse is not necessarily true.
}
\begin{equation}
\| \hat{a}(\bm{r}) \| \equiv \mathrm{Tr} | \hat{a}(\bm{r}) | = \Theta(1).
\label{eq:norm_a}
\end{equation}
Since $\hat{A} = O(N)$, 
two values of $\hat{A}$ are `distinct' if they are 
different by $\Theta(N)$.
According to this observation, the following definition is reasonable.
\begin{breakbox}\noindent
  {{\sf Defeinition:} Macroscopically distinct states}
  
Two (or more) states are {\em macroscopically distinct} if
the values of some additive observable $\hat{A}$ are different by $\Theta(N)$
between them.
\end{breakbox}

\subsection{Superposition of macroscopically distinct states}

Let $| A \nu \rangle$ be an eigenvector
of $\hat{A}$ corresponding to eigenvalue $A$, 
where $\nu$ labels degenerate eigenvectors.
If $|\psi \rangle$ does {\em not} contain a superposition of states 
with macroscopically distinct values of $A$, 
i.e., 
if it is just a superposition of $| A \nu \rangle$'s 
with macroscopically {\em non}-distinct values of $A$, 
then $|\langle A \nu |\psi \rangle|^2$ takes significant values 
only for $A$ such that 
$\left| A - \langle \psi | \hat{A} |\psi \rangle \right| = o(N)$.
In this case, 
$\langle \psi | (\Delta \hat{A})^2 |\psi \rangle = o(N^2)$.
Hence, by contradiction, 
if $\langle \psi | (\Delta \hat{A})^2 |\psi \rangle \neq o(N^2)$,
i.e.,
if $\langle \psi | (\Delta \hat{A})^2 |\psi \rangle = \Theta(N^2)$,
then 
$|\psi \rangle$ contains a superposition of states 
with macroscopically distinct values of $A$.
We are thus led to the following definition:
\begin{breakbox}\noindent{{\sf Definition:} (for pure states)
    Superposition of macroscopically distinct states}
  
A pure state $| \psi \rangle$ contains 
a {\em superposition of macroscopically distinct states}
if there exists an additive observable $\hat{A}$ such that 
$\langle \psi | (\Delta \hat{A})^2| \psi \rangle = \Theta(N^2)$.
\end{breakbox}

\subsection{Index $p$}

It is then convenient to define the index $p$ as follows.
\begin{breakbox}\noindent{{\sf Definition:} Index $p$ for pure states}
  
Let $\hat{A}$ be an additive observable.
For a pure state $| \psi \rangle$, the index $p$ is defined 
as a real number such that 
\begin{equation}
\max_{\hat{A}} \langle \psi | (\Delta \hat{A})^2| \psi \rangle = \Theta(N^p).
\end{equation}
\end{breakbox}
It is easy to show that $1 \leq p \leq 2$.
Using this index, the above definition can be rephrased as follows:
\begin{breakbox}\noindent{{\sf Definition:} Pure cat state
    
}
For a pure state $| \psi \rangle$, 
if $p=2$ then $| \psi \rangle$ contains 
a superposition of macroscopically distinct states,
which we call a {\em pure cat state}.
\end{breakbox}
Note that we have {\em never} assumed that 
only {\em two} macroscopically distinct states are superposed to form 
$|\psi \rangle$ with $p=2$.
Therefore, 
a pure cat state contains 
a superposition of {\em two or more} macroscopically distinct states.

For example, 
the cat state $| \mbox{cat} + \rangle$ has $p=2$, as expected, because $\langle (\Delta \hat{M_z})^2 \rangle = O(N^2)$. 
By contrast, 
$| \psi_1 \rangle$ has $p=1$, and hence is {\em not} a  pure cat state.
On the other hand, 
$| \psi_2 \rangle$ has $p=2$ because $\langle (\Delta \hat{M_z})^2 \rangle = \Theta(N^2)$, and hence is a  pure cat state.
This may be understood because $| \psi_2 \rangle$ is 
a superposition of states with 
 $M_z=\Theta(N)$ and $M_z=-\Theta(N)$.

Note that a state with 
$p = 2 - \epsilon$ ($0<\epsilon \ll 1$)
is close to, but not completely, 
a  pure cat state.
In this paper, we are not interested in such an 
incomplete superposition of macroscopically distinct states.
%

It was shown that 
$p$ is directly related to physics.
For example, 
fundamental stabilities of quantum many-body states are determined by $p$,
as described in Appendixes \ref{sec:sup:decoherence.p=2} and \ref{sec:sup:slm.p=2} 
and Refs.~\cite{shmzmiyadera2002,sugitashmz2005}.
Furthermore, $p=2$ is necessary for quantum computational speedup \cite{ukena2004appearance,speedup}.
It is known that index $p$ agrees with other measures for superposition of macroscopically distinct states \cite{shmzmiyadera2002,q,frowis2012measures,park2016quantum}.
These facts also support that $p$ is a reasonable index.
Furthermore, there is an efficient method of computing $p$ for a given pure state, as described in Appendix \ref{sec:sup:VCM} and Ref.~\cite{p}.
The reader, if not interested in these facts, may jump to 
Appendix \ref{sec:sup:q}, in which the index $q$ is reviewed.


\subsection{Decoherence rate of states with $p=2$}
\label{sec:sup:decoherence.p=2}

Let us consider the decoherence rate $\Gamma$ of a pure state 
$\ket{\psi}$ by a classical noise (or 
a perturbation from environments),
under a physical assumption that 
the interaction between the noise (or environments) and the system 
is the sum of local interactions.
It was shown in Refs.~\cite{shmzmiyadera2002,ukena2004appearance} that, with increasing the system size $N$, $\Gamma$
scales as
\begin{equation}
\Gamma \leq \Theta(N^p),
\label{eq:sup:Gamma}
\end{equation}
where $p$ is the index $p$ of $\ket{\psi}$. 
This is a {\em universal} result, independent of any details 
of the system and noise.
It implies, for example, that $\Gamma$ of a state with $p=1$ grows
at most as $\Theta(N)$.

It was also shown in Refs.~\cite{shmzmiyadera2002,ukena2004appearance} 
that the equality in Eq.~(\ref{eq:sup:Gamma}) is 
{\em achievable}, i.e., 
a noise achieving the equality is {\em in principle possible}\footnote{
We say ``in principle possible'' because 
we do not guarantee that such a noise really exists in a real physical 
system of interest.
} 
that satisfies the above assumption on interaction 
between the noise and the system.
In particular, if $\ket{\psi}$ has $p=2$
then a noise that satisfies the assumption 
is in principle possible such that 
$\ket{\psi}$ decoheres as fast as $\Gamma = \Theta(N^2)$.

\subsection{Stability against local measurements}
\label{sec:sup:slm.p=2}

Consider a quantum state $\hat{\rho}$, either pure or mixed,
which is translationally invariant.
Let $\hat{a}(x)$ and $\hat{b}(y)$ be local operators on 
(spatial regions around) the positions $x$ and $y$, 
respectively, which therefore commute with each other if $|x - y| >$
some constant.
By $P(a)$ [$P(b)$] we denote the probability of getting the outcome $a$ 
when $\hat{a}(x)$ [$\hat{b}(y)$] is measured.
By $P(a, b)$, we denote the probability of getting the outcome $a$ 
and $b$ when $\hat{a}(x)$ and $\hat{b}(y)$ are measured simultaneously.
We assume that 
$\hat{a}(x), \hat{b}(y)$ do not depend on $N$, and that
\begin{equation}
\braket{|\hat{a}(x)|}, 
\braket{|\hat{b}(y)|}
\leq \mbox{some constant independent of $N$}
\label{eq:sup:|a|}
\end{equation}
for all $\hat{a}(x), \hat{b}(y)$,  
where $\braket{\bullet}$ denotes the expectation value in $\hat{\rho}$.
 

In terms of these quantities, we here define the stability against local measurements
in a manner slightly different from that of Ref.~\cite{shmzmiyadera2002},
as follows\footnote{
While Ref.~\cite{shmzmiyadera2002} assumed natural states such as 
eigenstates of translationally invariant Hamiltonian with short-range
interactions, the result of present section is applicable to 
wider classes of states,
including artificial states which may appear in quantum information thoery,
if they are translationally invariant.
The cluster property (\ref{eq:sup:CP}) is 
defined also in a manner slightly different from that of Ref.~\cite{shmzmiyadera2002}.
}:

\begin{breakbox}\noindent{{\sf Definition:} Stability against local measurements}
  
We say $\hat{\rho}$ has {\em stability against local measurements}
if for any $\epsilon>0$ there exists $\ell_\epsilon$ such that (s.t.)
\begin{eqnarray}
&\left| P(a,b) - P(a)P(b) \right| 
\leq
\epsilon P(a)P(b)
\mbox{ for } ^\forall x, y \mbox{ s.t. }
|x-y| > \ell_\epsilon
\mbox{ and for } ^\forall \hat{a}(x), \hat{b}(y), a, b.
\label{eq:sup:stability}
\end{eqnarray}
\end{breakbox}
This implies that 
the outcome of measurement of $\hat{a}(x)$ at $x$ does not 
correlate with 
the outcome of measurement of $\hat{b}(y)$ at $y$ 
if $|x-y|$ is large enough.
When $P(b)>0$, we can rewrite (\ref{eq:sup:stability}) as
\begin{eqnarray}
\left| P(a;b) - P(a) \right| 
\leq
\epsilon P(a)
\mbox{ for } ^\forall x, y \mbox{ s.t. } |x-y| > \ell_\epsilon
\mbox{ and for } ^\forall \hat{a}(x), \hat{b}(y), a, b.
\label{eq:sup:stability2}
\end{eqnarray}
Here, 
$P(a;b) := P(a,b)/P(b)$ is the conditional probability
of getting $a$ by measurement of $\hat{a}(x)$ 
when 
$b$ is obtained by measurement of $\hat{b}(y)$.
Relation (\ref{eq:sup:stability2}) implies that 
measurement of a local observable does not affect 
the outcome of measurement of a local observable 
at a distant point.
We therefore call this property
stability against local measurements.

While this stability is expected for any stable macroscopic states,
it can be shown that 
a pure state with $p=2$ does {\em not} have this stability.

More generally, it can be shown that 
(\ref{eq:sup:stability2}) is equivalent to the ``cluster property'' 
in the following sense:
\begin{breakbox}\noindent{{\sf Definition:} Cluster property}
  
We say $\hat{\rho}$ has the {\em cluster property}
if for any $\epsilon>0$ there exists $\ell_\epsilon$ s.t.
\begin{eqnarray}
&\left| \braket{
\Delta \hat{a}(x) \Delta \hat{b}(y)
}\right| 
\leq
\epsilon \braket{|\hat{a}(x)|} \braket{|\hat{b}(y)|}
\mbox{ for } ^\forall x, y \mbox{ s.t. }
|x-y| > \ell_\epsilon
\mbox{ and for } ^\forall \hat{a}(x), \hat{b}(y).
\label{eq:sup:CP}
\end{eqnarray}
Here, 
$\Delta \hat{a}(x) := \hat{a}(x) - \braket{\hat{a}(x)}$
and 
$\Delta \hat{b}(y) := \hat{b}(y) - \braket{\hat{b}(y)}$.
\end{breakbox}
This 
is a generalization of the cluster property of infinite systems to finite systems.
For any pure state with $p=2$, we can show that 
it does not have the cluster property, 
and therefore it is not stable against local measurements.


\subsection{Variance-covariance matrix and off-diagonal long-range order}
\label{sec:sup:VCM}

There is an efficient method of calculating $p$ \cite{p}.
For simplicity, we assume that 
each site of the lattice is a spin-$1/2$ system. 
For a given pure state $|\psi \rangle$,
we define the variance-covariance matrix (VCM) by
\begin{align}
V_{\alpha i, \beta j} 
& :=
\langle\psi|
\Delta \hat{\sigma}_{\alpha}^i \Delta \hat{\sigma}_{\beta}^j
|\psi\rangle
\nonumber\\
& =
\langle\psi|
\hat{\sigma}_{\alpha}^i \hat{\sigma}_{\beta}^j
|\psi\rangle
-
\langle\psi| \hat{\sigma}_{\alpha}^i |\psi\rangle
\langle\psi| \hat{\sigma}_{\beta}^j |\psi\rangle,
\label{VCM}
\end{align}
where $\alpha,\beta=x,y,z$, and 
$i,j=1,2,\cdots,N$.
The VCM is 
a $3N\times 3N$ Hermitian non-negative matrix. 
If $e_{\rm max}$ is the maximum eigenvalue 
of the VCM, it is shown that 
\begin{equation}
e_{\rm max} = \Theta(N^{p-1}).
\end{equation}
One therefore has only to evaluate $e_{\rm max}$
to calculate $p$.
If $e_{\rm max} = \Theta(N)$, then $p=2$
and $|\psi\rangle$ contains superposition of 
macroscopically distinct values of 
the additive observable that is obtained from 
the eigenvector of the VCM corresponding to $e_{\rm max}$
\cite{p,morimaeshmz2006}.
It is also shown that 
the number of eigenvalues that scales as $\Theta(N)$ is at most 
$\Theta(1)$ \cite{morimaeshmz2006}.

Equation (\ref{VCM}) also show that the off-diagonal long-range order
does not necessarily imply $p=2$ \cite{shmzmiyadera2000,shmzmiyadera2002}.

\section{Index $q$ for mixed states}
\label{sec:sup:q}

When $\hat{\rho}$ is a mixed state,
$\mathrm{Tr} [ \hat{\rho} (\Delta \hat{A})^2 ] = \Theta(N^2)$
does not necessarily imply the existence of a 
superposition of macroscopically distinct states,
because such an equality is satisfied also for a (classical) {\em mixture} of macroscopically distinct states.
Hence, the index for mixed states cannot be a trivial generalization of $p$. 
To correctly identify superposition of macroscopically distinct states
 for mixed states,
the index $q$ was proposed in Ref.~\cite{q}, which we review in this section.

We assume that $\hat{\rho}$ is macroscopically uniform spatially. 

\subsection{Motivation}\label{sec:sup:motivation_q}

Consider the mixture
\begin{equation}
\hat \rho_{\rm ex1} 
\equiv 
{1 \over N} \sum_{i=1}^N | \psi_i \rangle \langle \psi_i |
\label{eq:sup:rhoex1}
\end{equation}
of $N$ different cat states,
\begin{equation}
| \psi_i \rangle
=
{1 \over \sqrt{2}}
\left( \ket{\bm{0}_i} + \ket{\bm{1}_i} \right)
\qquad (i = 1, 2, \cdots, N),
\label{eq:sup:psii}
\end{equation}
where 
\begin{align}
\ket{\bm{0}_i}
&:= | 0 \rangle^{\otimes (i-1)} | 1 \rangle | 0 \rangle^{\otimes (N-i)},
\\
\ket{\bm{1}_i}
&:= 
| 1 \rangle^{\otimes (i-1)} | 0 \rangle | 1 \rangle^{\otimes (N-i)}.
\end{align}
Every $| \psi_i \rangle$ is a cat state because 
$\ket{\bm{0}_i}$ and $\ket{\bm{1}_i}$ are 
eigenvectors of $\hat{M}_z$ 
with macroscopically distinct eigenvalues $\pm (N-2)$,
\begin{equation}
\hat{M}_z \ket{\bm{0}_i} = -(N-2) \ket{\bm{0}_i},
\quad
\hat{M}_z \ket{\bm{1}_i} = +(N-2) \ket{\bm{1}_i}.
\end{equation}
However, 
since the weight of each $| \psi_i \rangle$ 
in $\hat \rho_{\rm ex1}$ is as small as $1/N$,
which vanishes as $N \to \infty$, 
it may be nontrivial whether 
$\hat \rho_{\rm ex1}$
contains a 
superposition of macroscopically distinct states.

To inspect whether the superposition is contained, 
let us introduce the following witness observable:
\begin{equation}
\hat{W} :=
\sum_{i=1}^N 
\left( \ket{\bm{0}_i}
\bra{\bm{1}_i}
+
\ket{\bm{1}_i}
\bra{\bm{0}_i} \right),
\label{eq:sup:Wex1}
\end{equation}
whose eigenvalues are $0, \pm 1$.
Obviously, it can detect quantum coherence between 
$\ket{\bm{0}_i}$ and $\ket{\bm{1}_i}$, 
states with macroscopically distinct values of $\hat{M}_z$.
By noting
$
\braket{\psi_i | \hat{W} | \psi_i} = 1
$,
we find
\begin{equation}
\mathrm{Tr} [ \hat \rho_{\rm ex1} \hat{W} ]
=1,
\end{equation}
which shows that 
$\hat \rho_{\rm ex1}$
does contain superposition of macroscopically distinct states.
This may be understood by noting that 
$\hat \rho_{\rm ex1}$
 is a mixture of 
the `same sort' of superpositions of macroscopically distinct states
in the sense that 
all $| \psi_i \rangle$'s 
are superpositions of states with $M_z = \pm(N-2)$.

In this particular case, it was easy to guess the witness observable $\hat{W}$.
For general mixed states, however, it will be difficult to find 
an appropriate $\hat{W}$.
The idea of the index $q$ is to do this automatically.

\subsection{Index $q$}

Let 
$\hat{A}$ be an additive observable, 
$\hat{\eta}$ be a projection operator, and
\begin{equation}
\hat{C}_{\hat{A}, \hat{\eta}}
\equiv[\hat{A}, [\hat{A}, \hat{\eta}]],
\end{equation}
which is a correlation of local observables 
(see the example of the text). 
The index $q$ is defined by
\begin{breakbox}\noindent{{\sf Definition:} Index $q$ for mixed states}
  
For a mixed state $\hat{\rho}$, the index $q$ is defined as a 
real number such that 
\begin{equation}
\max \big\{
\max_{\hat{A}, \hat{\eta}}
\mathrm{Tr} \big( \hat{\rho} \hat{C}_{\hat{A}, \hat{\eta}} \big),
N \big\}
= \Theta(N^q).
\label{def:q:eta}
\end{equation}
\end{breakbox}
We can show that $1 \leq q \leq 2$.
Since $\hat{C}_{\hat{A}, \hat{\eta}}$ is a 
traceless Hermitian operator,
its eigenvalues are real numbers whose sum is $0$.
Using this property, the above definition can be rewritten as
\begin{equation}
\max \big\{
\max_{\hat{A}} 
\mathrm{Tr} \big| [\hat{A}, [\hat{A}, \hat{\rho}]] \big|, 
N \big\}
= \Theta(N^q).
\label{def:q:rho}
\end{equation}
Equation (\ref{def:q:eta}) is convenient when considering experiments
because 
\begin{equation}
\langle \hat{C}_{\hat{A}, \hat{\eta}} \rangle
\equiv
\mathrm{Tr} \big( \hat{\rho} \hat{C}_{\hat{A}, \hat{\eta}} \big)
\end{equation}
is the expectation value of the observable $\hat{C}_{\hat{A}, \hat{\eta}}$,
whereas 
(\ref{def:q:rho}) is often useful when studying theoretical aspects.

To see a physical meaning of $q$, we express 
$\langle \hat{C}_{\hat{A}, \hat{\eta}} \rangle$ 
as
\begin{equation}
\langle \hat{C}_{\hat{A}, \hat{\eta}} \rangle
=
\sum_j 
\sum_{A \nu A' \nu'}
(A-A')^2 
\langle \phi_j | A \nu \rangle
\langle A \nu |\hat{\rho} | A' \nu' \rangle 
\langle A' \nu' | \phi_j \rangle,
\end{equation}
where $| A \nu \rangle$ and $| \phi_i \rangle$
are eigenvectors 
($\nu$ labels degenerate eigenvectors) 
of $\hat{A}$ and $\hat{\eta}$, respectively:
\begin{align}
&
\hat{A} | A \nu \rangle = A | A \nu \rangle
\qquad (\langle A \nu | A' \nu' \rangle = \delta_{AA'} \delta_{\nu \nu'}),
\\
&
\hat{\eta} = \sum_j 
| \phi_j \rangle \langle \phi_j |
\qquad (\langle \phi_i | \phi_j \rangle = \delta_{ij}).
\end{align}
Suppose that, for some $(A, \nu, A', \nu')$,
\begin{equation}
(A-A')^2 = \Theta(N^2)
\quad \mbox{and} \quad
\langle A \nu |\hat{\rho} | A' \nu' \rangle = \Theta(1), 
\label{eq:A-A'=N2}
\end{equation}
which means that $\rho$ contains 
a superposition of states with macroscopically distinct values of $\hat{A}$.
One can detect this superposition as 
$\langle \hat{C}_{\hat{A}, \hat{\eta}} \rangle = \Theta(N^2)$ 
by taking $| \phi_j \rangle$ in such a way that 
\begin{equation}
\langle \phi_j | A \nu \rangle
\langle A \nu |\hat{\rho} | A' \nu' \rangle 
\langle A' \nu' | \phi_j \rangle
>0.
\label{eq:positivephase}
\end{equation}
When (\ref{eq:A-A'=N2}) is satisfied also for another $(A, \nu, A', \nu')$,
one can take another $| \phi_{j'} \rangle$ in such a way that 
(\ref{eq:positivephase}) is also satisfied
for this combination of $(A, \nu, A', \nu', | \phi_{j'} \rangle)$.
Consequently, both terms give 
$\langle \hat{C}_{\hat{A}, \hat{\eta}} \rangle = \Theta(N^2)$ 
when $\eta$ is appropriately taken.
This is the basic idea of defining a generalized cat state by $q$.
(Actually, more general cases can be treated by $q$, 
as exemplified in the next subsection.)
That is, it is defined as follows:
\begin{breakbox}\noindent{{\sf Definition:} Generalized cat state}
  
For a mixed state $\hat{\rho}$, 
if $q=2$ then $\hat{\rho}$ contains 
superpositions of macroscopically distinct states
with a significant magnitude.
We call such a state, for which 
$\langle \hat{C}_{\hat{A}, \hat{\eta}} \rangle = \Theta(N^2)$,
a {\em generalized cat state of $\hat{A}$}.
\end{breakbox}
A state with 
$q = 2 - \epsilon$ ($0<\epsilon \ll 1$)
is close to, but not completely, 
a generalized cat state.
In this paper, we are not interested in such an 
incomplete superposition of macroscopically distinct states.

If one is interested only in states with $q>1$
(such as the generalized cat states), 
the definition (\ref{def:q:eta}) of $q$ reduces to a simpler one,
\begin{equation}
\max_{\hat{A}, \hat{\eta}}
\mathrm{Tr} \big( \hat{\rho} \hat{C}_{\hat{A}, \hat{\eta}} \big)
= \Theta(N^q).
\label{def:q:simple}
\end{equation}
This simplified form is used in the paper because 
only states with $q=2$ are analyzed.
Note that when studying all states, including those with $q=1$, 
the original definition (\ref{def:q:eta}) should be used 
because otherwise some of the reasonable properties in the next section
would be lost.


\subsection{Properties of index $q$}\label{sec:sup:prop_q}

The index $q$ has the following properties:
\begin{enumerate}
\item\label{itm:separable}
 $q=1$ for any separable state (i.e., mixture of product states).

\item For pure states,
\begin{align}
q=2 \ &\Leftrightarrow \ p=2
\quad (\mbox{hence, } q<2 \ \Leftrightarrow \ p<2).
\label{q=2.equiv.p=2}
\\
q=1 \ &\Rightarrow \ p=1, 
\quad p=1 \ \Rightarrow \ q \leq 1.5.
\end{align}

\item Mixing can decrease $q$.

For example, consider two cat states,
\begin{equation}
| \mbox{cat} \pm \rangle
\equiv
{1 \over \sqrt{2}} \ 
| 0 0 0 \cdots 0 \rangle
\pm 
{1 \over \sqrt{2}} \ 
| 1 1 1 \cdots 1 \rangle,
\end{equation}
which have $q=p=2$.
Their mixture 
\begin{eqnarray}
\hat \rho_{\rm ex2}
&\equiv&
{1 \over 2} \ | \mbox{cat}+ \rangle \langle \mbox{cat}+|
+
{1 \over 2} \ | \mbox{cat}- \rangle \langle \mbox{cat}-|
\nonumber\\
&=&
{1 \over 2} \ 
| 1 1 1 \cdots 1 \rangle
\langle 1 1 1 \cdots 1 |
+
{1 \over 2} \ 
| 0 0 0 \cdots 0 \rangle
\langle 0 0 0 \cdots 0 |\nonumber\\
\label{eq:ex1}
\end{eqnarray}
is a separable state, hence $q=1$ according to property \ref{itm:separable}.

\item Mixing does not increase $q$, i.e., 
\begin{equation}
\hat{\rho} = \sum_i \lambda_i \hat{\rho}_i
\ \Rightarrow \
q \leq \max_i \{ q_i \}.
\end{equation}
This is evident from the trivial inequality;
\begin{equation}
\max_{\hat{A}, \hat{\eta}} 
\mathrm{Tr} \big( \hat{\rho} \hat{C}_{\hat{A}, \hat{\eta}} \big)
\leq
\sum_i \lambda_i
\max_{\hat{A}_i, \hat{\eta}_i} 
\mathrm{Tr} \big( \hat{\rho}_i \hat{C}_{\hat{A}_i, \hat{\eta}_i} \big).
\end{equation}
This inequality also shows the following.

\item 
If $\hat{\rho}$ has $q=2$ there exists a state(s) with $q=2$ in 
every decomposition.
That is, when 
\begin{equation}
\hat{\rho} 
= \sum_i \lambda_i \hat{\rho}_i
= \sum_i \lambda'_i \hat{\rho}'_i
= \cdots,
\end{equation}
where $0 \leq \lambda_i \leq 1$ and $\sum_i \lambda_i = 1$ 
and similarly for $\lambda'_i$,
then there exists a state with $q=2$ in each of 
$\{ \hat{\rho}_i \}_i$, $\{ \hat{\rho}'_i \}_i$, $\cdots$.

\item 
In particular, 
if $\hat{\rho}$ has $q=2$ there exists a pure state(s) with $p=2$ 
[which means $q=2$ according to (\ref{q=2.equiv.p=2})]
in every pure-states decomposition.
That is, when 
\begin{equation}
\hat{\rho} 
= \sum_i \lambda_i | \psi_i \rangle \langle \psi_i |
= \sum_i \lambda'_i | \psi'_i \rangle \langle \psi'_i |
= \cdots,
\end{equation}
where $0 \leq \lambda_i \leq 1$ and $\sum_i \lambda_i = 1$ 
and similarly for $\lambda'_i$,
then there exists a pure state with $p=2$ in each of 
$\{ | \psi_i \rangle \}_i$, $\{ | \psi'_i \rangle \}_i$, $\cdots$.
Here, the pure states in the decomposition are not necessarily
orthogonal to each other; e.g., 
we do {\em not} assume 
$\langle \psi_i | \psi_j \rangle =0$ for $i \neq j$.

\item In every pure-state decomposition, 
if $\hat{\rho}$ has $q=2$ then pure states with $p=2$ 
should be contained with a significant weight, i.e., 
\[
\sum_{i \ \in \ p=2} \lambda_i = \Theta(1).
\]
This is a {\em necessary} condition for $q=2$.

\item\label{itm:sufficient}
A {\em sufficient} condition for $q=2$ is as follows.
For an additive operator $\hat A$, suppose that 
pure states $| \psi_1 \rangle$, $| \psi_2 \rangle$, $\cdots$
satisfy
\begin{eqnarray}
&&
\langle \psi_i | \psi_{j} \rangle = \delta_{i, j}
\mbox{ for } i, j = 1, 2, \cdots,
\label{eq:cond1}\\
&&
\langle \psi_i | \hat A | \psi_{j} \rangle = 0
\mbox{ for } i \neq j,
\label{eq:cond2}\\
&&
\langle \psi_i | (\Delta_i \hat A)^2 | \psi_i \rangle
= \Theta(N^2) 
\mbox{ for } i \leq \Lambda,
\label{eq:cond3}\\
&&
\langle \psi_i | (\Delta_i \hat A)^2 | \psi_i \rangle
< \Theta(N^2) 
\mbox{ for } i > \Lambda,
\label{eq:cond4}\end{eqnarray}
where 
$\Delta_i \hat A \equiv
\hat A - \langle \psi_i | \hat A | \psi_i \rangle$
and $\Lambda$ is a positive integer.
Consider a classical mixture of these states,
$\hat \rho 
= \sum_i \lambda_i | \psi_i \rangle \langle \psi_i |$,
where $0 \leq \lambda_i \leq 1$ and $\sum_i \lambda_i = 1$.
If 
\begin{equation}
\sum_{i \leq \Lambda} \lambda_i = \Theta(1),
\label{eq:cond5}\end{equation}
then any such a mixture has $q=2$.
\end{enumerate}

\subsection{Examples}

In the example of Appendix \ref{sec:sup:motivation_q}, 
$| \psi_i \rangle$ of (\ref{eq:sup:psii}) satisfies
all conditions (\ref{eq:cond1})-(\ref{eq:cond3}) 
for $\hat A = \hat M_z = \sum_{\bm r} \hat \sigma_z({\bm r})$ 
and $\Lambda = N$.
Therefore, according to property \ref{itm:sufficient} of Appendix \ref{sec:sup:prop_q}, 
any mixtures of these states, such as 
$\hat \rho_{\rm ex1}$ of (\ref{eq:sup:rhoex1}),
have $q=2$.
This is consistent with the result on $\hat \rho_{\rm ex1}$ 
in Appendix \ref{sec:sup:motivation_q}, where  
we used the witness observable $\hat{W}$ that is explicitly given by
(\ref{eq:sup:Wex1}).
By using $q$, we have obtained the same conclusion without 
using an explicit form of a witness observable.

Another instructive example is the case where
\begin{equation}
| \varphi_i \rangle 
\equiv (| i \rangle +|\bar{i}\rangle)/ {\sqrt{2}},
\end{equation}
where $|i\rangle$ ($|\bar{i}\rangle$) 
is an arbitrary state in which $i$ spins are up (down) and 
$N-i$ spins are down (up).
If we limit the range of $i$ over, say, 
$1 \leq i \leq N/3$, then
conditions (\ref{eq:cond1})-(\ref{eq:cond3}) are all satisfied 
for $\hat A = \hat M_z$ and $\Lambda = N/3$.
Therefore, any mixtures of these states, such as
\begin{equation}
\hat \rho_{\rm ex3} \equiv 
{3 \over N} \sum_{i=1}^{N/3} 
| \varphi_i \rangle \langle \varphi_i |,
\end{equation}
have $q=2$.
Intuitively, 
such mixtures are mixtures of 
the same sort of superpositions of macroscopically distinct states
in the sense that 
all $| \varphi_i \rangle$'s 
are superpositions of states with positive and negative $M_z$.

Furthermore, consider mixtures of 
$\hat{\rho}_{\rm ex1}, \hat{\rho}_{\rm ex3}$ and 
$\hat{\rho}_{\rm ex2}$ of (\ref{eq:ex1}):
\begin{align}
\hat \rho'_{\rm ex1} &\equiv w \hat \rho_{\rm ex1} + (1-w) \hat \rho_{\rm ex2},
\\
\hat \rho'_{\rm ex3} &\equiv w \hat \rho_{\rm ex3} + (1-w) \hat \rho_{\rm ex2}.
\end{align}
They also have $q=2$ if $w > 0$ and independent of $N$,
because conditions (\ref{eq:cond1})-(\ref{eq:cond3}) are all satisfied.
This may be understood because they contain states with $q=2$ with 
significant weights.

These properties and examples show that 
$q$ is a reasonable index for a generalized cat state.
One can identify such a state by measuring 
$\hat{C}_{\hat{A}, \hat{\eta}}$ 
(correlation of local observables)
for an appropriate 
pair of $\hat{A}, \hat{\eta}$; 
hence, the state tomography is unnecessary.

\subsection{Quantifying superpositions of macroscopically distinct 
  states}
\label{sec:quantify}

The index $q$ is defined as the power of $N$ in 
$\braket{\hat C_{\hat A, \hat{\eta}}}$ because 
for large $N$ the power is,  obviously, more important than the coefficient. 
But, if necessary, one can use the value of 
$\braket{\hat C_{\hat A, \hat{\eta}}}$
to quantify superpositions of macroscopically distinct 
states, as follows\footnote{
For pure states, one can use the value of $\braket{(\Delta \hat{A})^2}$ 
to quantify superpositions of macroscopically distinct states, in a similar manner.}:

Consider two quantum states 
that are generalized cat states of the same\footnote{
To make quantitative comparison between 
generalized cat states of different additive observables, 
the normalization 
(\ref{eq:norm_a}) should be replaced with 
$|| \hat{a}(\bm{r}) || \equiv \mathrm{Tr} | \hat{a}(\bm{r}) | = Q$, a constant 
common to all local observable $\hat{a}(\bm{r})$.
} 
additive observable $\hat{A}$, say $\hat{M}_x$.
Let $\ket{m}$ satisfy $\hat M_x\ket{m}=m\ket{m}$.
Then, 
a pure state
$
\hat \rho_{N}:=\frac{1}{2}(\ket{N}+\ket{-N})(\bra{N}+\bra{-N}) 
$
gives $\braket{\hat{C}_{\hat{M}_x, \hat{\eta}}}=2N^2$ 
when $\hat \eta=\hat{\rho}_{N}$ (which maximizes $\braket{\hat{C}_{\hat{M}_x, \hat{\eta}}}$).
And
another pure state
$\hat \rho_{N/2}:=\frac{1}{2}(\ket{N/2}+\ket{-N/2})(\bra{N/2}+\bra{-N/2})$ gives $\braket{\hat{C}_{\hat{M}_x, \hat{\eta}}}=N^2/2$ 
when $\hat \eta=\hat{\rho}_{N/2}$ (which maximizes $\braket{\hat{C}_{\hat{M}_x, \hat{\eta}}}$).
From these observations, we see that $\braket{\hat{C}_{\hat{M}_x, \hat{\eta}}}$ quantifies 
how distinct the values of $\hat{M}_x$
between the states that are superposed in pure cat states.
As the difference of the eigenvalues ($m$'s) of superposed state becomes larger, $\braket{\hat{C}_{\hat{M}_x, \hat{\eta}}}$ also becomes larger.

On the other hand,
a mixed state
$\frac{1}{2}\hat \rho_{N}+\frac{1}{2}\hat \rho_{\mathrm{non-cat}}$ 
gives 
$\braket{\hat{C}_{\hat{M}_x, \hat{\eta}}}=N^2 +O(N)$
when $\hat \eta=\rho_{N}$ (which maximizes the leading-order term, i.e., the $O(N^2)$ term, of $\braket{\hat{C}_{\hat{M}_x, \hat{\eta}}}$).
Hence, we also see that $\braket{\hat{C}_{\hat{M}_x, \hat{\eta}}}$ 
decreases with decreasing the ratio of 
pure cat states in the mixed state.

{As seen from these examples, 
$\braket{\hat{C}_{\hat{M}_x, \hat{\eta}}}$
quantifies superpositions of macroscopically distinct 
states, reflecting the above two factors.


Applying this idea to our generalized cat state for free spins,
we see that the factor $\tanh^2 \beta h$ plays the essential role, quantifying the contribution from pure cat states. 
The ratio of 
the generalized cat state at $T>0$ 
and 
the generalized cat state at $T=0$, which is a pure cat state,
is
 $[2N+(N^2-M^2)\tanh^2 \beta h]/[2N+N^2-M^2]$ which behaves $\sim \tanh^2 \beta h$ for large $N$. 
Therefore, the factor $\tanh^2 \beta h$ reflects the decrease of the contribution of pure cat states when increasing the temperature.
}

\section{Purity of the post-measurement state}
For free spins, the post-measurement state is
\begin{align}
\hat \rho^0_{\mathrm{post}}
=
\frac{
\hat P_z \hat \rho_0^{\mathrm{eq}}  \hat P_z
}{
\mathrm{Tr}[\hat P_z \hat \rho_0^{\mathrm{eq}} \hat P_z]
}
=
\frac{
\hat P_z \hat e^{-\beta \hat H_0}  \hat P_z
}{
Z^0_{\mathrm{post}}(\beta h)
}.
\end{align}
Purity of this state is calculated as
\begin{align}
\mathrm{Tr}[(\hat \rho^0_{\mathrm{post}})^2]
&=
\frac{
\mathrm{Tr}[
\hat P_z \hat e^{-\beta \hat H_0} \hat P_z\hat e^{-\beta \hat H_0}\hat P_z
]}{
Z_0^2(\beta h)
}\\
&\leq
\frac{
\mathrm{Tr}[
\hat P_z (\hat e^{-\beta \hat H_0})^2 \hat P_z
]}{
Z_0^2(\beta h)
}\\
&=
\frac{Z^0_{\mathrm{post}}(2\beta h)}{Z^0_{\mathrm{post}}(\beta h)^2}.\label{upbound}
\end{align}
Since 
\begin{align}
Z^0_{\mathrm{post}}(\beta h)=\binom{N}{(N+M)/2}\cosh^N(\beta h),
\end{align}
the right-hand side of (\ref{upbound}) is calculated as
\begin{align}
&2^N\frac{
(e^{2\beta h}+e^{-2\beta h})^N
}{
\binom{N}{(N+M)/2}(e^{\beta h}+e^{-\beta h})^{2N}
}\nonumber\\
&=
\frac{
2^N
}{
\binom{N}{(N+M)/2}[1+2/(e^{2\beta h}+e^{-2\beta h})]^N
}\\
&=
\frac{
2^N
}{
\binom{N}{(N+M)/2}[1+1/\cosh(\beta h)]^N
}.
\end{align}
Using the Stirling formula,
we obtain
$\frac{
2^N
}{
\binom{N}{(N+M)/2}
}
\simeq
\sqrt{\frac{\pi N}{2}}\exp(M^2/2N)
$
when $|M|\lesssim \sqrt{N}$.
Hence when $|M|\lesssim \sqrt{N}$ and $\beta h=\Theta(1)$ (independent of $N$),
\begin{align}
\mathrm{Tr}[(\hat \rho^0_{\mathrm{post}})^2]
\leq
\frac{1}{e^{\Theta (N)}}.
\end{align}
Thus the post-measurement state is a mixture of 
$e^{\Theta (N)}$
states.
This is due to the properties of the pre-measurement state and the 
projection operator $\hat P_z$.
The pre-measurement state $\hat \rho_0^{\mathrm{eq}}$ is a mixture of 
$e^{\Theta (N)}$
states because its entropy
$ 
-\mathrm{Tr}[
\hat \rho_0^{\mathrm{eq}} \ln \rho_0^{\mathrm{eq}}
]
$ 
 is $\Theta(N)$ when $T>0$.
$\hat P_z$ is a projection onto $e^{\Theta (N)}$ dimensional space because
\begin{align}
\mathrm{Tr}[\hat P_z]=\binom{N}{(N+M)/2}\sim 2^N 
\end{align}
when $M\sim 0$.

When  $\beta h \rightarrow \infty$,
on the other hand, 
 the pre-measurement state is a ground state, i.e., a pure state.
In this case,
the post-measurement state is also a pure state.

\section{Calculation of energy for free spins}

Here we calculate $\braket{\hat H_0}_{\mathrm{post}}$ and $\braket{(\Delta\hat H_0)^2}_{\mathrm{post}}$ for the post-measurement state.
Using $\hat \sigma_x \ket{\uparrow}=\ket{\downarrow}$ and $\hat \sigma_x \ket{\downarrow}=\ket{\uparrow}$, we can calculate $\braket{\hat H_0}_{\mathrm{post}}=\mathrm{Tr}\left[\hat \rho^0_{\mathrm{post}}\hat H_0\right]$ easily as
\begin{align}
Z^0_{\mathrm{post}}(\beta,M)\braket{\hat H_0}_{\mathrm{post}} 
&= \mathrm{Tr}\left[\hat P_ze^{-\beta \hat{H}_0}\hat P_z\hat H_0\right]\\
&= -h\sum_{\xi}\braket{M,\xi|e^{-\beta \hat{H}_0}\hat P_z\hat{M}_x|M,\xi}\\
&= 0,
\label{1}
\end{align}
where $\hat M_z\ket{M,\xi}=M\ket{M,\xi}$ and 
$\xi$ labels degenerate eigenstates.
The last line comes from the fact that $\hat{M}_x\ket{M,\xi}=\sum_{i=1}^N\hat{\sigma}_x^i\ket{M,\xi}$ is a sum of the states that differ from $\ket{M,\xi}$ by one spin being flipped. Such states are also eigenstates of $\hat{M}_z$, but their eigenvalues are not $M$. Therefore, $\hat P_z\hat{M}_x\ket{M,\xi}=0$.

Similarly, using
\begin{align}
e^{-\beta\hat{H}_0}
=\left(
    \begin{array}{cc}
      \cosh\left(\beta h\right) & \sinh\left(\beta h\right) \\
      \sinh\left(\beta h\right) & \cosh\left(\beta h\right) 
    \end{array}
  \right)^{\otimes N}
\end{align}
and
\begin{align}
\braket{\uparrow|
\left(
    \begin{array}{cc}
      \cosh\left(\beta h\right) & \sinh\left(\beta h\right) \\
      \sinh\left(\beta h\right) & \cosh\left(\beta h\right) 
    \end{array}
  \right)
|\downarrow}
=
\braket{\downarrow|
\left(
    \begin{array}{cc}
      \cosh\left(\beta h\right) & \sinh\left(\beta h\right) \\
      \sinh\left(\beta h\right) & \cosh\left(\beta h\right) 
    \end{array}
  \right)
|\uparrow}
=
\sinh(\beta h),
\end{align}
$Z^0_{\mathrm{post}}(\beta,M)\braket{\hat H_0^2}_{\mathrm{post}}$ is calculated as
 \begin{align}
   Z^0_{\mathrm{post}}(\beta,M) \braket{\hat H_0^2}_{\mathrm{post}}
   &= \mathrm{Tr}\left[\hat P_ze^{-\beta \hat{H}_0}\hat P_z\hat H_0^2\right]\\
&= h^2\sum_{\xi}\braket{M,\xi|e^{-\beta \hat{H}_0}\hat P_z\hat{M}_x^2|M,\xi}\\
&= h^2Z^0_{\mathrm{post}}(\beta,M)\left(N+\frac{N^2-M^2}{2}\tanh^2(\beta h)\right)\label{2}.
\end{align}
Thus we have 
\begin{align}
\braket{(\Delta \hat H_0)^2}_{\mathrm{post}}
&=\braket{\hat H_0^2}_{\mathrm{post}}-\braket{\hat H_0}_{\mathrm{post}}^2 \\
&= h^2\left(N+\frac{N^2-M^2}{2}\tanh^2(\beta h)\right)\\
&= \Theta(N^2),
\end{align}
which assures us $q=2$ when
\begin{align}
N^2-M^2=\Theta(N^2). 
\end{align}  

\section{Calculation for the $XYZ$ model}\label{sec:xyz}

For $\hat H = \hat H_0+\hat H_{\mathrm{int}}=-h\hat M_x -\sum_{i=1}^N\left(
J_x\hat{\sigma}^i_x\hat{\sigma}^{i+1}_x+
J_y\hat{\sigma}^i_y\hat{\sigma}^{i+1}_y+
J_z\hat{\sigma}^i_z\hat{\sigma}^{i+1}_z
\right) $, the `partition function' and the density matrix of the post-measurement state are 
\begin{align}
Z_{\mathrm{post}}&:=\mathrm{Tr}\left[\hat P_ze^{-\beta \hat{H}}\hat P_z\right],\\
\hat{\rho}_{\mathrm{post}}&:=\frac{\hat P_ze^{-\beta \hat{H}}\hat P_z}{Z_{\mathrm{post}}}.
\end{align}

According to the discussion on the general systems and states in Sec.~VII, $\braket{\hat C}_{\mathrm{post}}=\Theta(N^2)$ for interacting spins. Let us take a look at the coefficients of the $\Theta(N^2)$ term.

\subsection{Calculation of $Z_{\mathrm{post}}$}
We are going to calculate $Z_{\mathrm{post}}$ up to $\beta^2$ order.
 This will give a good result when 
\begin{align}
|\beta \boldsymbol{J}|\ll 1\label{smallj}
\end{align}
and 
\begin{align}
|\beta h|\ll1 ,\label{smallh}
\end{align}
where $\boldsymbol{J}=(J_x,J_y,J_z)$.
Using the notation
\begin{align}
\hat{K}_{\alpha}:=-J_{\alpha}\sum_{i=1}^N\hat{\sigma}^i_{\alpha}\hat{\sigma}^{i+1}_{\alpha}
\end{align}
for ${\alpha}=x,y,z$, we can expand $e^{-\beta \hat{H}}$ as
\begin{align}
e^{-\beta \hat{H}}&= 1-\beta \hat{H}+\frac{\beta^2}{2}\hat{H}^2+O(\beta^3).
\end{align}
Substituting this into the definition of $Z_{\mathrm{post}}$ and dropping the terms that are obviously zero,
 we have
\begin{align}
Z_{\mathrm{post}}
&=\sum_{\xi}\braket{M,\xi|e^{-\beta \hat{H}}|M,\xi}\\
&=\sum_{\xi}\bra{M,\xi}1-\beta \hat{K}_z
  +\frac{\beta^2}{2}\left(
\hat{K}_x^2+\hat{K}_y^2+\hat{K}_z^2+\hat{H}_0^2-2J_xJ_y\sum_i\hat{\sigma}_z^i\hat{\sigma}_z^{i+1}
\right)\ket{M,\xi}
+O(\beta^3)\\
&=\binom{N}{\frac{N+M}{2}}\left(
1 + \frac{\beta^2}{2}N(J_x^2+J_y^2+h^2)
\right)
-(\beta J_z +\beta^2J_xJ_y)\sum_{\xi}\sum_i\bra{M,\xi}\hat{\sigma}_z^i\hat{\sigma}_z^{i+1}\ket{M,\xi}\nonumber\\
&+\frac{\beta^2 J_z^2}{2}\sum_{\xi}\sum_{i,j}\braket{M,\xi|\hat{\sigma}_z^i\hat{\sigma}_z^{i+1}\hat{\sigma}_z^j\hat{\sigma}_z^{j+1}|M,\xi}
+O(\beta^3).\label{ztotyuu}
\end{align}
After some algebra, we have
\begin{align}
Z_{\mathrm{post}}
&=\binom{N}{\frac{N-M}{2}}\left(
1 + \frac{\beta^2}{2}N(J_x^2+J_y^2+h^2)
-(\beta J_z +\beta^2J_xJ_y)\frac{M^2-N}{N-1}\nonumber\right.\\
&\left.
+\frac{\beta^2 J_z^2
\left(
-2M^4(N-4)+(N-3)(N-2)N(N(N+4)+16)
+M^2(N(N(N-15)+90))-112
\right)
}{
2(M-N-2)(N-2)(N-1)(N+M+2)
}
\right) \label{indepz}\\
&=\binom{N}{\frac{N-M}{2}}\left(
1-\beta J_z\frac{M^2-N}{N-1}+O(\beta^2)
\right).
\end{align}
As we will see later, details of the $O(\beta^2)$ terms of $Z_{\mathrm{post}}$ will not be necessary for the purpose of knowing $\mathrm{Tr}\left[
\hat{\rho}_{\mathrm{post}}\hat{C}_{\hat{M}_x,\hat P_z}
\right]$ up to $\beta^2$ order.

\subsection{Index $q$ of $\hat{\rho}_{\mathrm{post}}$}
To evaluate
$\mathrm{Tr}[\hat{\rho}_{\mathrm{post}} \hat{C}_{\hat{M}_x,\hat P_z}]
=2\mathrm{Tr}[\hat{\rho}_{\mathrm{post}} \hat M_x^2]$,
 we evaluate $\mathrm{Tr}\left[\hat P_ze^{-\beta \hat{H}}\hat P_z\hat{M}_x^2\right]$.
Assuming (\ref{smallj}) and (\ref{smallh}), we obtain
\begin{align}
&\mathrm{Tr}\left[\hat P_ze^{-\beta \hat{H}}\hat P_z\hat{M}_x^2\right]\nonumber\\
&=NZ_{\mathrm{post}}
+2\binom{N}{\frac{N-M}{2}}\left(
-\beta J_x \frac{N^2-M^2}{2(N-1)}-\beta J_y \frac{N^2-M^2}{2(N-1)}\right.\nonumber\\
&\left.
+\beta^2J_x^2\frac{N^2-M^2}{2(N-1)}+\beta^2J_y^2\frac{N^2-M^2}{2(N-1)}
+\beta^2h^2\frac{N^2-M^2}{4}\right.\nonumber\\
&\left.+\beta^2J_z(J_x+J_y)\frac{(N^2-M^2)(M^2-2N+4)}{2(N-1)(N-2)}
\right).
\end{align}
With this and $Z_{\mathrm{post}}$ from the previous subsection, $\braket{\hat{C}_{\hat{M}_x,\hat P_z}}_{\mathrm{post}}$ for the post-measurement state is obtained as
\begin{align}
\braket{\hat{C}_{\hat{M}_x,\hat P_z}}_{\mathrm{post}}
&=2\mathrm{Tr}\left[\hat P_ze^{-\beta \hat{H}}\hat P_z\hat{M}_x^2\right]/Z_{\mathrm{post}}\\
&=2N
-2\beta \frac{N^2-M^2}{N-1}(J_x+J_y)\nonumber\\
&+2\beta^2(N^2-M^2)\left(
\frac{h^2}{2}+\frac{J_x^2+J_y^2}{N-1}\right.\nonumber\\
&\left.+\frac{J_z(J_x+J_y)(M^2-N^2+4N-4)}{(N-1)^2(N-2)}
\right)
+O(\beta^3)\\
&=2N+
\beta O(N)+
\beta^2h^2(N^2-M^2)+
\beta^2 O(N)
+O(\beta^3)\label{intyuragi}.
\end{align}
The third term in the right-hand side indicates that up to the order of $\beta^2$, the coefficient of the $\Theta(N^2)$ term does not depend on $\boldsymbol{J}$.
    
When $J_y=J_z$  ($=:J_\perp$), we can improve (\ref{intyuragi}) by expanding
 only $\exp(-\beta \hat H_{\mathrm{int}})$ since 
$
e^{-\beta \hat H}=\exp(-\beta \hat H_0)\exp(-\beta \hat H_{\mathrm{int}}).
$
Calculating in the same manner,
we obtain
\begin{align}
\braket{\hat{C}_{\hat{M}_x,\hat P_z}}_{\mathrm{post}}
&=
2N+(N^2-M^2)\tanh^2(\beta h)
-2\beta(J_x+J_{\perp})\frac{N^2-M^2}{N-1}\nonumber\\
&+2\beta^2(N^2-M^2)\left(
\frac{J_x^2}{N-1}+\frac{J_xJ_{\perp}(M^2-N^2+4N-4)}{(N-1)^2(N-2)}\right.\nonumber\\
&\left.+\frac{J_{\perp}(M^2+N-2)}{(N-1)^2(N-2)}
\right)+O(\beta^3).
\end{align}

The second term in the right-hand side indicates that up to the order of $\beta^2$, the coefficient of the $\Theta(N^2)$ term is the same for the free spins.
This result is useful when a strong magnetic field is applied, i.e., $|\beta h|\gg 1$, while the interactions between spins are weak, i.e., $|\beta \boldsymbol{J}|\ll 1$.


\section{Symmetry consideration}\label{sp:sym}

By revisiting the $XYZ$ model from the symmetry point of view, we find that the discussion on the coefficient of $N^2$ is applicable to a broader class of Hamiltonian.
To show it, we denote the rotation around $z$ axis by angle $\pi$, by $\hat R_z$.
Since 
\begin{align}
\hat R_z\hat H_0\hat R_z^{\dag}
=-\hat H_0,\\
\hat R_z\hat H_{\mathrm{int}}\hat R_z^{\dag}
=\hat H_{\mathrm{int}},\\
\hat R_z\hat P_z\hat R_z^{\dag}
=\hat P_z,
\end{align}
 we find that $Z_{\mathrm{post}}$ and $\braket{\hat C_{\hat M_x \hat P_z}}_{\mathrm{post}}$ are even functions of $\beta h$, i.e.,
\begin{align}
&Z_{\mathrm{post}}(\beta h,\beta \boldsymbol{J},M)\nonumber\\
&=\mathrm{Tr}[\hat P_z e^{-\beta(\hat H_0+\hat H_{\mathrm{int}})}\hat P_z]\\
&=\mathrm{Tr}[\hat R_z\hat P_z \hat R_z^{\dag}\hat R_ze^{-\beta(\hat H_0+\hat H_{\mathrm{int}})}\hat R_z^{\dag}\hat R_z\hat P_z\hat R_z^{\dag}]\\
&=Z_{\mathrm{post}}(-\beta h,\beta \boldsymbol{J},M),\\
&C(\beta h,\beta \boldsymbol{J},M)\nonumber\\
&:=\braket{\hat C_{\hat M_x \hat P_z}}_{\mathrm{post}}\\
&=2\mathrm{Tr}[\hat P_z e^{-\beta(\hat H_0+\hat H_{\mathrm{int}})}\hat P_z\hat M_x^2/Z_{\mathrm{post}}(\beta h,\beta \boldsymbol{J},M)]\\
&=\frac{2\mathrm{Tr}[\hat R_z\hat P_z \hat R_z^{\dag}\hat R_ze^{-\beta(\hat H_0+\hat H_{\mathrm{int}})}\hat R_z^{\dag}\hat R_z\hat P_z\hat R_z^{\dag}\hat R_z\hat M_x^2\hat R_z^{\dag}]}{Z_{\mathrm{post}}(-\beta h,\beta \boldsymbol{J},M)}\\
&=C(-\beta h,\beta \boldsymbol{J},M)
\end{align}
This implies that, if we expand 
$
C(\beta h, \beta \boldsymbol{J}, M)
$
in a power series of $\beta h$ and $\beta \boldsymbol{J}$, then 
\begin{align}
&C(\beta h, \beta \boldsymbol{J}, M)\nonumber\\
&=C^{(0)}(\beta h, \beta \boldsymbol{J}, M)
+C^{(1)}(\beta h, \beta \boldsymbol{J}, M)\nonumber\\
&+C^{(2)}(\beta h, \beta \boldsymbol{J}, M)
+\cdots\\
&=C^{(0)}(0,0, M)
+C^{(1)}(0, \beta \boldsymbol{J}, M)
+C^{(2)}(0, \beta \boldsymbol{J}, M)\nonumber\\
&+C^{(2)}(\beta h, 0, M)
+\cdots.  
\end{align}
According to the result for the free spins,
 $C(\beta h, 0, M)=(N^2-M^2)\tanh^2(\beta h)$, and thus
 $C^{(2)}(\beta h, 0, M)=(N^2-M^2)(\beta h)^2$, which indicates that up to to the order of $\beta^2$, the coefficient of $N^2$ is the same as the free spins for any dimension and any lattice.

We immediately see that we can extend this discussion to a more general Hamiltonian.
That is, if a system has the Hamiltonian $\hat H_{z-\mathrm{inv}}$ which is invariant under $\hat R_z$, 
then, regardless of the dimension nor the details of the lattice, the coefficient of $N^2$ is the same as the free spins for $O(\beta^2)$.
Note that the above discussion does not restrict the range of interaction between spins. Hence it is applicable, e.g.,  even to the systems with long-range interactions.

\section{Projection onto a finite interval}\label{sec:appdebu}
\subsection{Case for free spins}
We consider the free spins $\hat H_0$ and the projection operator $\hat P'_z$ onto the subspace corresponding to an interval $\alpha_-N \leq \hat M_z \leq \alpha_+N$.
We assume $|\alpha_-|<\alpha_+$ without loss of generality.
In this case, the `partition function' is
\begin{align}
Z'^0_{\mathrm{post}}
&:=\mathrm{Tr}\big[\hat P'_ze^{-\beta \hat{H}_0}\hat P'_z\big]\\
&=\sum_{\ket{M,\xi}}\braket{M,\xi|\hat P'_ze^{-\beta \hat{H}_0}\hat P'_z|M,\xi}\\
&=\sum_{\ket{M,\xi}\,s.t.\,\alpha_-N\leq M \leq \alpha_+N}\braket{M,\xi|e^{-\beta\hat{H}_0}|M,\xi}\\
&=r(\alpha_-N,\alpha_+N)\cosh^N\left(\beta h\right),
\end{align}
where $\ket{M,\xi}$ is an eigenstate of $\hat M_z$, $\xi$ labels degenerate eigenstates, and
\begin{align}
r(\alpha_-N,\alpha_+N)
:=\sum_{k=0}^{\frac{\alpha_+N-\alpha_-N}{2}}\binom{N}{\frac{N+\alpha_-N}{2}+k}.\label{r}
\end{align}
Then the post-measurement state is given by 
$\hat \rho'^0_{\mathrm{post}}:=\hat P'_ze^{-\beta \hat{H}_0}\hat P'_z/Z'^0_{\mathrm{post}}$.
Taking $\hat A=\hat M_x$ and $\hat \eta=\hat P'_z$ for $\braket{\hat C_{\hat A,\hat \eta}}_{\mathrm{post}}$ for $\hat \rho'^0_{\mathrm{post}}$, we have
\begin{align}
&\braket{\hat C_{\hat M_x, \hat P'_z}}_{\mathrm{post}}\nonumber\\
&=\mathrm{Tr}\left[
\hat \rho'^0_{\mathrm{post}}\left(
\hat M_x^2\hat P'_z-2\hat M_x\hat P'_z\hat M_x+\hat P'_z\hat M_x^2
\right)
\right]\nonumber\\
&=
\mathrm{Tr}\left[
\frac{\hat P'_ze^{-\beta \hat{H}_0}\hat P'_z}{Z'^0_{\mathrm{post}}}\left(
\hat M_x^2\hat P'_z-2\hat M_x\hat P'_z\hat M_x+\hat P'_z\hat M_x^2
\right)
\right]\\
&=\sum_{\ket{M,\xi}\,s.t.\,\alpha_-N\leq M \leq \alpha_+N}
\frac{2
\braket{M,\xi|
e^{-\beta \hat{H}_0}\hat P'_z\hat M_x\left(\hat{1}-\hat P'_z\right)\hat M_x
|M,\xi}}{Z'^0_{\mathrm{post}}}.\label{debuneko}
\end{align}
To calculate the right-hand side of this, we count the number of $\ket{M,\xi}$'s that go out of $\left[\alpha_-,\alpha_+\right]$ by operation of the first $\hat M_x$ [reading (\ref{debuneko}) from right to left], and come back by the second $\hat M_x$.
 Then we find
\begin{align}
&\sum_{\ket{M,\xi}\,s.t.\,\alpha_-N\leq M \leq \alpha_+N}
\frac{2
\braket{M,\xi|
    e^{-\beta \hat{H}_0}\hat P'_z\hat
    M_x\left(\hat{1}-\hat P'_z\right)\hat M_x
    |M,\xi}}{Z'^0_{\mathrm{post}}}\nonumber\\ 
&=\frac{2}{Z'^0_{\mathrm{post}}}\left( \cosh^N(\beta
  h)\left(\binom{N}{\frac{N-\alpha_+N}{2}}\frac{N-\alpha_+N}{2}
+\binom{N}{\frac{N+\alpha_-N}{2}}\frac{N+\alpha_-N}{2}\right)\right.\nonumber\\ &\left.+\sinh^2(\beta
  h)\cosh^{N-2}(\beta
  h)\left(\binom{N}{\frac{N-\alpha_+N}{2}}\frac{N-\alpha_+N}{2}\frac{N+\alpha_+N}{2}\right.\right.\nonumber\\
&\left.\left.+\binom{N}{\frac{N+\alpha_-N}{2}}\frac{N+\alpha_-N}{2}\frac{N-\alpha_-N}{2}\right)
  \right)\\ 
&=N^2\tanh^2(\beta h)I(N,\alpha_+N,\alpha_-N)
+O(N),
\end{align}
where
\begin{eqnarray}
I(N,\alpha_+N,\alpha_-N)
:=
\frac{r(\alpha_+N,\alpha_+N)(1-\alpha_+^2)}{2r(\alpha_-N,\alpha_+N)}
 +\frac{r(\alpha_-N,\alpha_-N)(1-\alpha_-^2)}{2r(\alpha_-N,\alpha_+N)}.
\end{eqnarray}
The right-hand side of Eq.~(\ref{debuc}) 
becomes $\Theta (N^2)$ if 
$I(N,\alpha_+N,\alpha_-N)$,
which is by definition between $0$ and $1$,
 is $\Theta (1)$.
Thus we investigate the conditions for $\alpha_+N$ and $\alpha_-N$ to satisfy it.
Since $r(\alpha_-N,\alpha_-N)>r(\alpha_+N,\alpha_+N)$,
it is necessary and sufficient for 
$\braket{\hat C_{\hat M_x \hat P'_z}}=\Theta (N^2)$
that $
r(\alpha_-N,\alpha_-N)(1-\alpha_-^2)/r(\alpha_-N,\alpha_+N)=\Theta (1)
$.
We define $g(x)$ as follows:
\begin{align}
g(x)&:=r(xN,xN)=\binom{N}{\frac{1+x}{2}N}
\end{align}
Using the Stirling formula, we have
\begin{align}
g(x)
&\sim \frac{1}{\sqrt{\pi (1-x^2)N/2}}
\left(\frac{2}{1+x}\right)^{\frac{1+x}{2}N}
\left(\frac{2}{1-x}\right)^{\frac{1-x}{2}N}\\
&=\frac{1}{\sqrt{\pi (1-x^2)N/2}}
\exp\left[
N\left(
\frac{1+x}{2}\ln \frac{2}{1+x}
+\frac{1-x}{2}\ln \frac{2}{1-x}
\right)
\right].
\end{align}
When $x \ll 1$, in particular, 
\begin{align}
g(x)\simeq \frac{2^N}{\sqrt{\pi N/2}}e^{-\frac{N-1}{2}x^2}.
\end{align}
We find that $g(x)$ is convex up when $x<1/\sqrt{N-1}\sim 1/\sqrt N$, and convex down when $x>1/\sqrt{N-1}\sim 1/\sqrt N$.
Using these, we can obtain the following conditions for obtaining a generalized cat state.
\begin{itemize}
\item When $\alpha_->1/\sqrt N$ and $\alpha_- \rightarrow 0$ ($N \rightarrow \infty$),
$\alpha_+-\alpha_-\leq \Theta(1/N)$ is the necessary and sufficient condition for $r(\alpha_-N,\alpha_-N)/r(\alpha_-N,\alpha_+N)=\Theta(1)$.

\item When $\alpha_+< 1/\sqrt N$ and $\alpha_->0$,
convergence of $N(\alpha_+-\alpha_-)$ to a positive constant
is the necessary and sufficient condition.

\item When $0<\alpha_- \leq 1/\sqrt N \leq \alpha_+$,
$\alpha_+-\alpha_-= \Theta(1/N)$ is the necessary and sufficient condition.

\item Using the results we have obtained, we can show that 
$\alpha_+ \leq \Theta(1/N)$ is the necessary and sufficient condition
when $\alpha_- \leq 0 \leq \alpha_+$.

\item When $\alpha_-=\Theta(1)$ and $1-\alpha_-=\Theta(1)$, 
generalized cat state can be obtained for any $\alpha_+$.
\end{itemize}

\subsection{Case for interacting spins}

It is easy to show that the discussion on $\hat P'_z$ holds the same for the $XYZ$ model.
Using the symmetry
\begin{align}
\hat R_z \hat P'_z\hat R_z^{\dag}=\hat P'_z,
\end{align}
we find that 
\begin{align}
Z'_{\mathrm{post}}
:=\mathrm{Tr}\big[\hat P'_ze^{-\beta (\hat{H}_0 +\hat H_{\mathrm{int}})}\hat P'_z\big]
\end{align}
and
\begin{align}
\hat \rho'_{\mathrm{post}}
:=\mathrm{Tr}\big[\hat P'_ze^{-\beta (\hat{H}_0 +\hat H_{\mathrm{int}})}\hat P'_z\big]/Z'_{\mathrm{post}}
\end{align}
are even functions of $\beta h$,
since $\hat R_z\hat H_0\hat R_z^{\dag}=-\hat H_0$ and $\hat R_z\hat H_{\mathrm{int}}\hat R_z^{\dag}=\hat H_{\mathrm{int}}$.
Following the discussion in Appendix \ref{sp:sym} (with $\hat P_z$ replaced with $\hat P_z'$),
we expect that if the conditions for $q=2$ for $\hat H_0$ are satisfied, 
then $q=2$ for $\hat H_0+\hat H_{\mathrm{int}}$.

\section{Feasibility in an experiment}\label{sec:appfeasi}

A single 
NV$^-$ center in diamond is known to have a long coherence time such as 
$\tau=470$ $\mu$s at room temperature \cite{balasubramanian2009ultralong,ishikawa2012optical,maurer2012room}.
Let us consider a system composed of $N$ NV$^-$ centers.

Let $N(t)$ be the number of NV$^-$ centers that maintain coherence at time $t$.
When decoherence occurs independently in individual NV$^-$ centers, 
\begin{align}
N(t)=N\exp(-t/\tau).
\end{align}
We are interested in how long {\em all} the spins maintain coherence,
which defines the coherence time $\tau_{\rm coh}$ of this system.
Thus we solve $N(\tau_{\rm coh})=N-1$.
When $N\gg 1$, we obtain
\begin{align}
\tau_{\rm coh}=\tau/N.
\end{align}
Therefore, $\tau_{\rm coh}=4.7$ $\mu$s and $47$ ns
for $N=100$ and $N=10^4$, respectively.
Taking $3.5$ $\mu$s ($35$ ns) as a duration of the measurement for $N=100$ ($N=10^4$),
we estimate how close the SERF magnetometer \cite{dang2010ultrahigh} should be 
to detect a magnetic field by one spin.

The SERF magnetometer realized by a group at Princeton \cite{dang2010ultrahigh} has 
magnetic field sensitivity $\delta B=160$ aT/$\sqrt {\mathrm{Hz}}$.
This means,
if the measurement lasts for $3\tau_{\rm coh}/4\simeq3.5$ $\mu$s,
the magnetometer can detect 
 the magnetic field with 
\begin{align}
(160\times 10^{-18}\,\mathrm{T}/\sqrt {\mathrm{Hz}})\times \left(\frac{1} {2\times 3.5 \times 10^{-6} \,\mathrm{s}}\right)^{1/2}
=60 \,\mathrm{fT}\label{est}
\end{align}
sensitivity. 
($600$ fT if $35$ ns.)

One spin has a magnetic moment of $\mu_{B}=9.3\times 10^{-24} \,\mathrm{A}\,\mathrm{m^2}$.
In a vacuum,
the magnetic field $B$ made by the spin is, when the distance from the spin is $r$,
\begin{align}
B=\frac{\mu_{B}  \mu_0}{2\pi r^3},
\end{align}
where $\mu_{0}=1.2\times10^{-6} \,\mathrm{T}\,\mathrm{m/A}.$
If $r=3\times10^{-6} \,\mathrm{m}=3 \,\mu\mathrm{m}$, then
\begin{align}
B=
65 \,\mathrm{fT}.
\end{align}
This is larger than the estimation of (\ref{est}).
Thus, 
for $N=100$,
$\hat M_z$ measurement of $\Theta(1)$ resolution is possible
if we assume that the outcome of the measurement is the time average of the measured magnetic field.
After the measurement, a generalized cat state should be obtained, and it survives for the rest of the system's coherence time, e.g., $\tau_{\rm coh}/4\simeq1.2$ $\mu$s.

For $N=10^4$, smaller $r$ is necessary since the coherent time $\tau_{\rm coh}$ of the system decreases. 
We find that $r=1$ $\mu$m is sufficient.
With $r=1$ $\mu$m, 
 the magnetic field which one spin makes is $B=1.8$ pT,
which is larger than $1.7$ pT, the minimum magnetic field the SERF can resolve within $\tau_{\rm coh}/10$ for $N=10^4$.
Then the obtained cat state survives for $9\tau_{\rm coh}/10$.
For $N=100$ and $r=1$ $\mu$m, we find that $\tau_{\rm coh}/10^3$ is sufficient for a measurement time, resulting in a long lifetime of the obtained cat state.

As discussed in \ref{sec:sup:decoherence.p=2},
decoherence rate {\em could be}, in principle, as large as $\Theta (N^2)$.
That is, coherence time of the obtained generalized cat state {\em could be} $t=\tau/N^2$.
When there is such a noise,        
the above estimation for $N=100$ is replaced with the case for $N=10^4$.
That is, for  $N=100$, 
distance of the magnetometer must be $r=1$ $\mu$m and the duration of the measurement must be $4.7$ ns.


\section{Verification of the generalized cat state in experiments}\label{sec:veri}

\subsection{Expression of $\hat C_{\hat M_x, \hat P_z}$ by the Pauli operators}
Here, as an illustration, we express ${\hat C_{\hat M_x \hat P_z}}$ of the free spins $\hat H_0$ of $N=4$ spins with the outcome of the measurement $M_z=0$ by the Pauli operators,
 since what are easily measured in experiments would be $\hat \sigma_{\alpha}$ ($\alpha=x,y,z$) of each spin.
Then projection operator $\hat P_z$ is
\begin{align}
\hat P_z&=
\left(
\frac{\sum_{l=1}^4\hat \sigma_z^l}{4}-\hat 1
\right)
\left(
\frac{\sum_{l=1}^4\hat \sigma_z^l}{4}+\hat 1
\right)
 \left(
\frac{\sum_{l=1}^4\hat \sigma_z^l}{2}-\hat 1
\right)
\left(
\frac{\sum_{l=1}^4\hat \sigma_z^l}{2}+\hat 1
\right)\\
&=
-\frac{1}{8}\sum_{l<m}\hat \sigma_z^l\hat \sigma_z^m
+\frac{3}{8}\prod_l\hat \sigma_z^l
+\frac{3}{8}\hat 1.
\end{align}
Then,
\begin{align}
{\hat C_{\hat M_x \hat P_z}}
&=[\hat M_x,[\hat M_x, \hat P_z]]\\
&=[\sum_{i=1}^N\hat \sigma_x^i,[\sum_{i=1}^N\hat \sigma_x^i,-\frac{1}{8}\sum_{l<m}\hat \sigma_z^l\hat \sigma_z^m
+\frac{3}{8}\prod_l\hat \sigma_z^l
+\frac{3}{8}\hat 1
]]\\
&=-\sum_{l<m}\left(
\hat \sigma_z^l\hat \sigma_z^m-\hat \sigma_y^l\hat \sigma_y^m
\right)\nonumber\\
&+3\big(
2\hat \sigma_z^1\hat \sigma_z^2\hat \sigma_z^3\hat \sigma_z^4 
-\hat \sigma_y^1\hat \sigma_y^2\hat \sigma_z^3\hat \sigma_z^4
-\hat \sigma_y^1\hat \sigma_z^2\hat \sigma_y^3\hat \sigma_z^4 
-\hat \sigma_y^1\hat \sigma_z^2\hat \sigma_z^3\hat \sigma_y^4
-\hat \sigma_z^1\hat \sigma_y^2\hat \sigma_y^3\hat \sigma_z^4 
-\hat \sigma_z^1\hat \sigma_y^2\hat \sigma_z^3\hat \sigma_y^4
-\hat \sigma_z^1\hat \sigma_z^2\hat \sigma_y^3\hat \sigma_y^4
\big).\label{ssss}
\end{align}
The term 
$\hat \sigma_z^1\hat \sigma_z^2\hat \sigma_z^3\hat \sigma_z^4$
can be measured, for example, by measuring 
$\hat \sigma_z^1, \hat \sigma_z^2, \hat \sigma_z^3$, 
and $\hat \sigma_z^4$ simultaneously.
This measurement also gives 
$\hat \sigma_z^l \hat \sigma_z^m$
for all pairs of $l, m$.
On the other hand,
 $\hat \sigma_y^i$ and $\hat \sigma_z^i$ cannot be measured simultaneously 
 (with vanishing errors)
because of their noncommutativity. 
Hence,  Eq.~(\ref{ssss}) implies that 
to measure $\hat C_{\hat M_x \hat P_z}$
(and thereby to confirm the success of the conversion into a generalized cat state)
for $N=4$  and $M=0$, 
it is sufficient to perform 
seven types of simultaneous measurement 
 corresponding to the last seven terms in Eq.~(\ref{ssss}).

\subsection{How many types of measurements 
are necessary for general $N$?}

We assume that $N$ is an even number.
Then,   
the projection operator onto $\hat M_z=M$ subspace is 
\begin{align}
\hat P_z:=
\prod_{L(\neq M)}\frac{\hat M_z-L}{M-L}.
\end{align}
Hence 
\begin{align}
{\hat C_{\hat M_x \hat P_z}}
&=[\hat M_x,[\hat M_x,\hat P_z]]\\
&=[\hat M_x,[\hat M_x,\sum_{l=0}^Na_l\hat M_z^l]]
\end{align}
for a certain set of $\{a_l\}$ ($l=1,2,...N$).
To find how many types of measurements are necessary, we look at $l=N$ term.
In particular, we investigate 
$[\hat M_x,[\hat M_x,\prod_{i=1}^N\hat \sigma_z^i]]$.
Using $[\hat \sigma_x,\hat \sigma_z]=-2i\hat \sigma_y$ and $[\hat \sigma_x,\hat \sigma_y]=2i\hat \sigma_z$, we find that 
$[\hat M_x,[\hat M_x,\prod_{i=1}^N\hat \sigma_z^i]]$ will give
$\prod_{i=1}^N\hat \sigma_z^i$ [one type], and products of $N-2$ $\hat \sigma_z$'s and two $\hat \sigma_y$'s [$\binom{N}{2}$ types].
Thus, one must perform at most $(N^2-N)/2+1$ types of simultaneous measurement.

\section{Pre-measurement state may be non-equilibrium}

\subsection{When pre-measurement state is $\hat P_x$}

We consider a quantum system with $N$ spins. Dimension of the Hilbert space is $D:=2^N$.
To a state
\begin{align}
\hat \rho:=\frac{1}{D}\hat 1,
\end{align}
we first operate a projection operator $\hat P_x:=\sum_{\nu}\ket{M_x,\nu}\bra{M_x,\nu}$ onto the $\hat M_x=M_x$ subspace,
 and then operate a projection operator $\hat P_z:=\sum_{\xi}\ket{M_z,\xi}\bra{M_z,\xi}$ onto the $\hat M_z=M_z$ subspace, 
where $\nu$ and $\xi$ label degenerate eigenstates of $\hat M_x$ and $\hat M_z$ respectively.

The state after operating $\hat P_x$ is 
\begin{align}
\hat \rho_{M_x}
&:=
\frac{
\hat P_x \hat \rho \hat P_x
}{
\mathrm{Tr}\left[\hat P_x \hat \rho \hat P_x\right]
}\\
&=\frac{
\hat P_x\hat 1 \hat P_x
}{
\mathrm{Tr}\left[\hat P_x\right]
}\\
&=\frac{
\hat P_x
}{
\binom{N}{\frac{N+M_x}{2}}
}.
\end{align}
To this $\rho_{M_x}$, $\hat P_z$ is operated. 
Then the state becomes
\begin{align}
&\hat \rho_{M_z}\nonumber\\
&:=\frac{
\hat P_z \hat P_x \hat P_z
}{
\mathrm{Tr}\left[\hat P_z \hat P_x \hat P_z\right]
}\\
&=\frac{
\sum_{\xi,\xi'}\ket{M_z,\xi}\bra{M_z,\xi}\left(\sum_{\nu}\ket{M_x,\nu}\bra{M_x,\nu}\right)\ket{M_z,\xi'}\bra{M_z,\xi'}
}{
\mathrm{Tr}\left[\hat P_z \hat P_x \hat P_z\right]
}.
\end{align}
Does this $\hat \rho_{M_z}$ have $q=2$? Taking $\hat A=\hat M_x$ and $\eta=\hat P_z$, we calculate $\braket{\hat C_{\hat A\hat \eta}}$ as
\begin{align}
&\mathrm{Tr}\left[\hat \rho_{M_z}\hat C_{\hat M_x \hat P_z}\right]\nonumber\\
&=\frac{
\mathrm{Tr}\left[
\hat P_z \hat P_x \hat P_z \left(\hat M_x^2\hat P_z-2\hat M_x\hat P_z \hat M_x +\hat P_z \hat M_x^2\right)
\right]
}{
\mathrm{Tr}\left[\hat P_z \hat P_x \hat P_z\right]
}\\
&=\frac{
2\mathrm{Tr}\left[\hat P_z \hat P_x \hat P_z \hat M_x^2\right]
}{
\mathrm{Tr}\left[\hat P_z \hat P_x \hat P_z\right]
}\\
&=
2N
+\frac{
4\sum_{\xi}\sum_{\ket{M_z,\xi}'}\bra{M_z,\xi}\hat P_x  \ket{M_z,\xi}'
}{
\mathrm{Tr}\left[\hat P_z \hat P_x \hat P_z\right]
},\label{c}
\end{align}
where $\ket{M_z,\xi}'$ is a state that differs from $\ket{M_z,\xi}$ by one $\ket{\uparrow}$ and one $\ket{\downarrow}$ being flipped.
After some algebra, we obtain
\begin{align}
\braket{\hat C}&=2N+
(N^2-M_z^2)\left(
1-\frac{N^2-M_x^2}{N(N-1)}
\right)\\
&=2N+
(N^2-M_z^2)\frac{M_x^2-N}{N(N-1)}.
\end{align}
Thus $\hat \rho_{M_z}$ is a generalized cat state when $(N^2-M_z^2)=\Theta(N^2)$ and $|M_x|=\Theta(N)$.

\subsection{$\braket{\hat M_x}_{\mathrm{pre}}=\Theta(N)$
is a sufficient condition}\label{sec:koashi}

More generally, the pre-measurement state $\hat \rho_{\mathrm{pre}}$ 
may be arbitrary if it has a macroscopic value of $\hat M_x$, 
i.e.\  if
$\braket{\hat M_x}_{\mathrm{pre}}=\Theta(N)$,
because then the conditions are satisfied
 by the set of 
$\{ \hat{M}_x$ or $\hat{M}_y, \hat{M}_z, \hat \rho_{\mathrm{pre}} \}$
for the following reason.
[We thank M. Koashi for suggesting this point.]

The probability of getting the outcome $\hat{M_z}=M$ is given by
\begin{equation}
{\rm Pr}(M)
=
\braket{\hat P_z}_{\mathrm{pre}},
\end{equation}
and 
the post-measurement state by
\begin{equation}
\hat P_z \hat \rho_{\mathrm{pre}} \hat P_z
/
{\rm Pr}(M).
\end{equation}
Instead of investigating 
$\braket{\hat{M}_x^2}_{\mathrm{post}}$
 and
$\braket{\hat{M}_y^2}_{\mathrm{post}}$
 separately,
we study
$\braket{\hat{M}_x^2 + \hat{M}_y^2}_{\mathrm{post}}$.
Furthermore, 
instead of investigating it for each value of $M$, 
we consider its average over $M$.
Since
 $[\hat M_x^2+M_y^2,\hat P_z]=0$,
it is evaluated as
\begin{align}
\sum_M 
{\rm Pr}(M)
\braket{\hat{M}_x^2 + \hat{M}_y^2}_{\mathrm{post}}
&=
\sum_M 
\mathrm{Tr}[\hat P_z\hat \rho_{\mathrm{pre}} 
(\hat M_x^2+\hat M_y^2)]
\nonumber\\ 
&=
\braket{\hat M_x^2+\hat M_y^2}_{\mathrm{pre}}
\nonumber\\ 
&\geq 
\braket{\hat M_x^2}_{\mathrm{pre}}
\nonumber\\ 
&\geq 
(\braket{\hat M_x}_{\mathrm{pre}})^2
\nonumber\\ 
&=\Theta(N^2).
\end{align}
This shows that a generalized cat is obtained with non-vanishing 
probability.
%

\section{Time evolution}\label{sec:time}
After the $\hat M_z$ measurement,
the system evolves autonomously with time.
In some of the models we have investigated, 
such as in the free spins, 
 $\hat M_x$ is conserved because $[\hat H,\hat M_x]=0$.
Since the post-measurement states are generalized cat states of $\hat M_x$,
the state will continue having $q=2$ for these models.
 In fact,
if $\hat M_z$ is measured at $t=0$
 its post-measurement state $\hat \rho_{\mathrm{post}}$ evolves as
 $
\hat U_t \hat \rho_{\mathrm{post}} \hat U^{\dag}_t
$, where $\hat U_t:=e^{-i\hat H t}$.
 Hence,
taking $\hat A=\hat M_x$ and $\hat \eta = \hat U_t \hat P_z \hat U^{\dag}_t=:\hat P(t)$ for $\hat C_{\hat A \hat \eta}$,
we have $\mathrm{Tr}\big[\hat U_t \hat \rho_{\mathrm{post}} \hat U^{\dag}_t \hat C_{\hat M_x \hat P(t)}\big]
=\mathrm{Tr}\big[\hat \rho_{\mathrm{post}} \hat C_{\hat M_x \hat P_z}\big]
=\Theta (N^2)$; thus $q=2$.
Note that taking $\hat \eta=\hat P(t)$ is equivalent to taking $\hat C_{\hat A \hat \eta}=\hat U_t \hat C_{\hat M_x \hat P_z}\hat U^{\dag}_t$ because $[\hat U_t, \hat M_x]=0$.
Therefore, to verify the generalized cat state,
one should measure the observable (\ref{ssss})
in which $\hat \sigma_y$ and $\hat \sigma_z$ are replaced with 
$\hat U\hat \sigma_y\hat U^{\dag}_t$ 
and $\hat U_t\hat \sigma_z\hat U^{\dag}_t$,  respectively 
[which are $\cos(2ht)\hat \sigma_y-\sin(2ht)\hat \sigma_z$ and $\sin(2ht)\hat \sigma_y+\cos(2ht)\hat \sigma_z$ 
for  free spins].

\end{widetext}

\bibliography{ref}

\end{document}